\documentclass[12pt]{article}
\pdfoutput=1

\usepackage{color}
\usepackage{amsmath}
\usepackage{amsfonts}
\usepackage{amssymb}
\usepackage{caption}
\usepackage{graphicx}
\usepackage{slashed}            % for slashed characters in math mode
\usepackage{subfig}             % for sub figures
\usepackage{xspace}				% For spacing after commands
\usepackage{cite}
\usepackage[english]{babel}
\usepackage{fancyhdr}
\usepackage{amsmath}
\usepackage{amssymb}
\usepackage{amsfonts}
\usepackage{psfrag}
\usepackage[applemac]{inputenc}
\usepackage{graphicx}
\usepackage{braket}
\usepackage{breqn}
\usepackage{feynmp}
\usepackage{feynmp-auto}
\usepackage{subfig}
\usepackage[export]{adjustbox}
\usepackage[toc,page]{appendix}
\usepackage{mathtools} 
\usepackage{float}
\usepackage[mathscr]{euscript}
\usepackage{breqn}
\usepackage{setspace}

\newcommand{\arXivold}[2]{\href{http://arxiv.org/pdf/#1}{{\tt #2/#1}}}

\usepackage[
	colorlinks=true,
	citecolor=black,
	linkcolor=black,
	urlcolor=blue,
	hypertexnames=false]{hyperref}

\numberwithin{equation}{section} 

\begin{document}
\begin{titlepage}

\begin{center}

	{	% TITLE GOES HERE 	
		\LARGE \bf 
		The Scale of New Physics from the \\ Higgs Couplings to $\gamma\gamma$ and $\gamma Z$
	}
	
\end{center}
	\vskip .3cm

\begin{center} % AUTHORS HERE
{\bf \  Fayez Abu-Ajamieh\footnote{\tt
		 \href{mailto:fayez.abu-ajamieh@umontpellier.fr}{fayezajamieh@IISc.ac.in}
\setcounter{footnote}{0}
		 }
		} 
\end{center}

\begin{center} 

 % PLACES HERE
	{Center for High Energy Physics (CHEP), Indian Institute of Science (IISc) \newline
	C.V. Raman Avenue, Bangalore 560012 - India}

\end{center}

%%%%%%%%%%%%%%%%%%%%%%%%%%%%%%%%%%%%%%%%%%%%%%%%%%%%%%%%%%%%

\centerline{\large\bf Abstract}

\begin{quote}
Measuring the Higgs couplings accurately at colliders is one of the best routes for finding physics Beyond the Standard Model (BSM). If the measured couplings deviate from the SM predictions, then this would give rise to energy-growing processes that violate tree-level unitarity at some energy scale, indicating new physics. In this paper, we extend previous work on unitarity bounds from the Higgs potential and the Higgs couplings to vector bosons and the top quark; to the Higgs couplings to $\gamma\gamma$ and $\gamma Z$. We find that while the HL-LHC might be able to find new physics in the $\gamma Z$ sector, the scale of new physics in both sectors is mostly beyond its reach. However, accurate measurements of the leading couplings of the two sectors in the HL-LHC can place stringent limits on both the scale of new physics and on other Higgs couplings that are difficult to measure. In addition, the scale of new physics is mostly within the reach of the $100$ TeV collider.

\end{quote}

\end{titlepage}

%%%%%%%%%%%%%%%%%%%%%%%%%%%%%%%%%%%%%%%%%%%%%%%%%%%%%%%%%%
%%%%%%%%%%%%%%%%%%%%%%%%%%%%%%%%%%%%%%%%%%%%%%%%%%%%%%%%%%

\section{Introduction}
\begin{figure}[!t]
\centerline{\begin{minipage}{0.8\textwidth}
\centering
\centerline{\includegraphics[width=400pt]{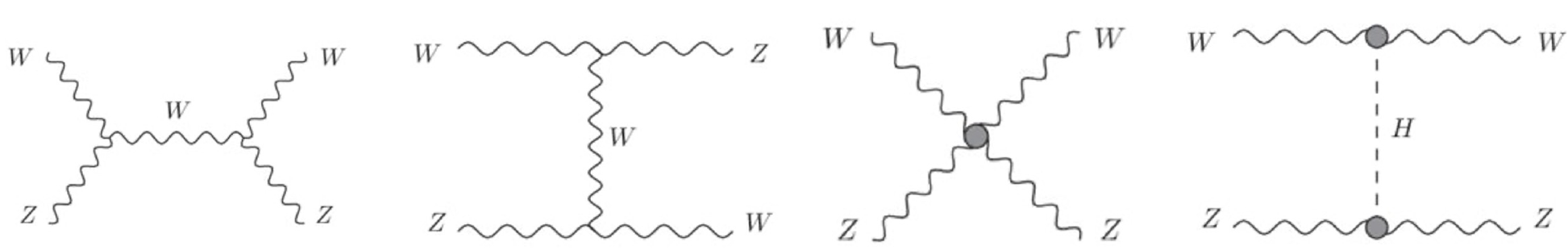}}
\caption{\small The $s$ and $u$ channels and the contact term for the scattering $W_{L} Z_{L} \rightarrow W_{L} Z_{L}$, in addition to the Higgs exchange diagram which unitarizes the former channels.}
\label{fig1}
\end{minipage}}
\end{figure}

Long before its discovery at the LHC in 2012, the need for the Higgs boson was firmly established based purely on unitarity arguments.\footnote{The other proposal to unitarize the electroweak sector is of course technicolor, which is now disproven.} It has long been known that the electroweak sector was incomplete without the Higgs boson, as the unitarity of amplitudes at high energy requires the theory to be spontaneously broken \cite{LlewellynSmith:1973yud, Cornwall:1973tb, Joglekar:1973hh, Cornwall:1974km} (see \cite{Aoude:2019tzn, Durieux:2019eor, Bachu:2019ehv} for a modern approach). Unitarity was used by Lee, Quigg, and Thacker in \cite{Lee:1977yc, Lee:1977eg} in order to put an upper bound on the Higgs mass of $\sim 1$ TeV. More specifically, consider the scattering $W_{L} Z_{L} \rightarrow W_{L} Z_{L}$; if one adds the amplitudes of the $s$ and $u$ channels together with the contact term (the first three diagrams in Figure \ref{fig1}), then one obtains an amplitude that grows quadratically with energy
\begin{equation}\label{eq:stuScatter}
\mathcal{M}_{s} + \mathcal{M}_{u} + \mathcal{M}_{\text{cont.}} = \frac{t}{v^{2}} + O(1) \sim \frac{E^{2}}{v^{2}} + \cdots.
\end{equation}

This quadratic growth of the amplitude with energy eventually violates unitarity at some scale, which signals the onset of new physics. On the other hand, adding the Higgs exchange diagram (the last diagram in Fig. \ref{fig1}) unitarizes the amplitude, as the amplitude gives
\begin{equation}\label{eq:HiggsEx}
\mathcal{M}_{H} = -\frac{t}{v^{2}} + O(1) \sim -\frac{E^{2}}{v^{2}} + \cdots,
\end{equation}
which exactly cancels the quadratic divergence in eq. (\ref{eq:stuScatter}). Requiring that unitarity be preserved translates into an upper bound on the Higgs mass
\begin{equation}\label{eq:HiggsUpperBound}
M_{H} \leq \Big(\frac{8\sqrt{2}\pi}{3G_{F}} \Big)^{\frac{1}{2}} \simeq 1 \hspace{1 mm} \text{TeV},
\end{equation}
and indeed, the Higgs was discovered in the LHC with a mass of $125$ GeV \cite{ATLAS:2012yve, CMS:2012qbp}. As a matter of fact, this argument was one of the main motivations for building the LHC.

In the same spirit, unitarity arguments can be utilized to set an upper limit on the scale of new physics. The reason is that the SM is the unique UV-complete theory with the observed particle content that can be extrapolated to arbitrarily high energies, which means that any deviation from the SM predictions will ruin this UV-completeness. This UV-incompleteness will manifest itself as energy-growing amplitudes that eventually violate unitarity at some high energy scale, signaling the onset of new physics, in exactly the same way as Lee, Quigg, and Thacker's argument points to the upper limit on the Higgs mass.

The Higgs couplings to other SM particles are only constrained at $O(\sim 20\%)$ at best, whereas its self-interaction is only measured within a factor of $\sim 10$. This leaves ample room for physics BSM to be probed at the High Luminosity (HL) LHC. This argument was used in \cite{Chang:2019vez, Abu-Ajamieh:2020yqi} to probe the scale of new physics in Higgs potential, as well as the Higgs couplings to the top quark and the $W$ and $Z$. 

In this study, we seek to extend this argument to the $h\gamma\gamma$ and $h \gamma Z$ sectors. We will analyze the $hgg$ sector in \cite{Abu-Ajamieh}. In the SM, the Higgs couples to $\gamma\gamma$ and $\gamma Z$ at loop level. In this paper, we employ an effective coupling approach where the loops are integrated out in order to write these couplings as tree-level couplings. The current level of experimental measurements constrains the coupling $h\gamma\gamma$ to be within $O(0.1)$, whereas the coupling $h\gamma Z$ is only constrained within a factor of $\sim 2$. This leaves plenty of room for physics BSM.

Previous studies in the literature utilized the SM Effective Field Theory (SMEFT) to investigate the couplings $h\gamma\gamma$ \cite{Hartmann:2015aia, Hartmann:2015oia} and $h\gamma Z$ \cite{Dawson:2018pyl}. However, in this paper, we adopt the more model-independent bottom-up approach that was used in \cite{Chang:2019vez, Abu-Ajamieh:2020yqi}, which does not rely on power counting, but parametrizes the scale of new physics as deviations from the SM predictions. We do not make any assumptions regarding the deviations and the infinitely many unconstrained higher-order couplings, other than that they are compatible with experimental measurements. This approach avoids the SMEFT assumption of only keeping the LO operator and neglecting higher-order ones. In addition, this approach makes it easier to generalize $2 \rightarrow 2$ scattering to $n \rightarrow m$ scattering, which gives stronger bounds.

As we show in this study, due to the loop suppression of the Higgs coupling to $\gamma\gamma$ and $\gamma Z$, the scale of new physics is generally much higher than what was obtained from the tree-level couplings in \cite{Chang:2019vez, Abu-Ajamieh:2020yqi}. For instance, it was found in \cite{Chang:2019vez, Abu-Ajamieh:2020yqi}, that the current level of constraints on the couplings $hVV$ and $h\bar{t}t$ allows for a scale of new physics that could be as low as $\sim 3$ TeV, which is well within the reach of the HL-LHC. On the other hand, we will show in this paper that the model-independent scale of new physics from the Higgs coupling to $\gamma\gamma$ and $\gamma Z$ is generally $\gtrsim 10$ TeV unless the deviations from the SM predictions are unnaturally large. This is beyond the reach of the HL-LHC, however, we show that one can probe these couplings indirectly, making these sectors worth investigating.

This paper is organized as follows: In Section \ref{Sec:Review}, we present a comprehensive review of the model-independent approach and the main results found in \cite{Chang:2019vez, Abu-Ajamieh:2020yqi}. In Sections \ref{Sec:hgg} and \ref{Sec:hgZ}, we investigate new physics in the $h \gamma\gamma$ and the $h\gamma Z$ sectors respectively, where we also discuss the relationship with SMEFT. We relegate much of the technical details, including defining the states and deriving the unitarity condition, to the appendices. We discussion the Electroweak Precision Observables (EWPO) in Section \ref{sec:EWPO}, and we present several simple UV completions in Section \ref{sec:UVcompletion}. Finally, we present our conclusions in Section \ref{sec:conclusions}.
%=============================================================================

% =============================================================================
\section{Review of the Model-independent Approach}\label{Sec:Review}
This section is intended as a comprehensive review of the model-independent approach utilized in \cite{Chang:2019vez, Abu-Ajamieh:2020yqi} and the main results therein so that the reader can obtain all salient points herein. Readers who are interested in further details regarding the scale of new physics from the Higgs potential, as well as from the Higgs interactions with the top quark and the $W/Z$, are instructed to refer to \cite{Chang:2019vez, Abu-Ajamieh:2020yqi}. 

In the literature, physics BSM is usually captured through the SMEFT approach, where one would systematically enumerate all higher-order operators that are consistent with the symmetries of the SM, suppressed by the appropriate power of the UV scale $\Lambda$. Another approach is the Higgs Effective Field Theory (HEFT), which is a different parameterization of the most general physics beyond the Standard Model. SMEFT and HEFT are only different by the power counting rule that determines the relative importance of the different contributions. That is to say, there is no universally correct power expansion, and different types of new physics have effective theories with different expansions.

While SMEFT (and to a lesser extent HEFT) is quite systematic and elegant, it is nonetheless not quite model-independent and relies upon several implicit assumptions, as a single UV scale of new physics is assumed in the operator expansion, which needn't be the case. For example, the scale suppressing derivative operators could be different from the scale suppressing non-derivatives ones, and one could envisage conspiratorial cancelations between the two types. Another possibility is that the new physics could come from heavy particles whose masses arise from electroweak symmetry breaking which do not decouple at large mass. In addition, when utilizing SMEFT, only the leading order SMEFT operators are commonly retained, which implicitly assumes that the scale of new physics is high enough that higher-order operators can be safely neglected, an assumption that might not hold for all we know. For example, if the scale of new physics is only a few TeV, then one can show that the dimension 8 SMEFT operators are comparable to the dimension 6 ones. Furthermore, SMEFT is not very transparent phenomenologically, as the couplings are what is measured in colliders and not the UV scale $\Lambda$. Finally, one might want to allow for the possibility of anomalously small coefficients due to accidental symmetries or conspiratorial cancelations, a possibility that could be missed by simple power expansion.

Given these considerations, we are thus motivated to adopt a different bottom-up approach, where the new physics in the Effective Field Theory (EFT) is captured through deviations in the Higgs couplings from the SM predictions. The only constraint on these deviations is that they are compatible with experimental measurements. Thus, we are maximally conservative and make no assumption with regard to the size of these deviations. These deviations could receive contributions from any number of higher-dimensional operators in an expansion, i.e., a single deviation could receive contribution from a tower of higher-order operators. As we show later on, there is a one-to-one correspondence between this approach and SMEFT, thereby making the lack of an expansion scale in this approach justified. With this approach, we can write the effective Lagrangian of the Higgs potential and its interactions with the top quark and massive gauge bosons in the unitary gauge as follows
\begin{flalign}\label{eq:effLag1}
\mathcal{L} & = \mathcal{L}_{\text{SM}} - \delta_{3} \frac{m_{h}^{2}}{2v}h^{3} - \delta_{4} \frac{m_{h}^{2}}{8v^{2}}h^{4} - \sum_{n=5}^{\infty}\frac{c_{n}}{n!}\frac{m_{h}^{2}}{v^{n-2}}h^{n} +\nonumber \cdots \\
& + \delta_{Z1}\frac{m_{Z}^{2}}{v}h Z_{\mu}Z^{\mu} +\delta_{W1}\frac{2m_{W}^{2}}{v}hW_{\mu}^{+}W^{\mu-} + \delta_{Z2}\frac{m_{Z}^{2}}{2v^{2}}h^{2} Z_{\mu}Z^{\mu} +\delta_{W2}\frac{m_{W}^{2}}{v^{2}}h^{2}W_{\mu}^{+}W^{\mu-}   \\
& + \sum_{n=3}^{\infty}\Big[ \frac{c_{Zn}}{n!}\frac{m_{Z}^{2}}{v^{n}}h^{n}Z_{\mu}Z^{\mu} + \frac{c_{Wn}}{n!}\frac{2m_{W}^{2}}{v^{n}}h^{n}W^{+}_{\mu}W^{-\mu} \Big] +\nonumber \cdots \\
& -\delta_{t1} \frac{m_{t}}{v}h \bar{t}t - \sum_{n=2}^{\infty}\frac{c_{tn}}{n!}\frac{m_{t}}{v^{n}}h^{n}\bar{t}t + \nonumber \cdots
\end{flalign}
where $\delta_{x}$ parametrize the deviations in the couplings $g_{x}$ compared to the SM predictions\footnote{Notice that in the $\kappa$ framework, $\kappa_{x} = 1 + \delta_{x}$}

\begin{equation}\label{eq:delta}
\delta_{x} \equiv \frac{g_{x} - g_{x}^{(\text{SM})}}{g_{x}^{(\text{SM})}},
\end{equation}
whereas $c_{i}$ represent the Wilson coefficients of higher-dimensional operators that do not have SM counterparts. Notice here that we have divided the operators by appropriate powers of $v$ to keep the Wilson coefficients dimensionless, i.e., the powers of $v$ are NOT an expansion scale and are only introduced for convenience. Therefore, $\delta_{i}$ and $c_{i}$ could assume any value compatible with experimental measurements. 

Before we proceed, we should clarify a few of points. In this approach, what we are actually doing is writing the higher-dimensional operators with effective couplings to perform simplified tree-level calculations, which is suitable for determining the scale of unitarity-violation. On the other hand, the lack of an expansion scale means that performing loop-level calculations with this approach would require special care as we shall see in Section \ref{sec:EWPO}. We also should note that one could have additional operators in eq. (\ref{eq:effLag1}) by introducing higher derivatives, however, higher derivatives will only lead to amplitudes that grow faster with energy and thus can only lower the unitarity-violating scale. We choose to be conservative and neglect operators with higher derivatives. Finally, we point out that although in this paper we call the results that depend only on one deviation model-independent, as they do not depend on any UV assumptions; we should emphasize that they nonetheless do depend on the basis chosen for the higher-dimensional operators. Consider $\delta_{3}$ for example; although there are processes that only depend on $\delta_{3}$, and are thus independent of other assumptions, one can nonetheless remove $\delta_{3}$ entirely by redefining the basis via introducing the correlated deviations in the operators $h\partial^{2}h$ and $hhVV$, which would corresponding to eliminating the operator $|H^{\dagger}H|^{3}$ in favor of $(\partial_{\mu}|H^{\dagger}H|^{2})^{2}$ and $|H^{\dagger}H||D_{\mu}H|^{2}$ in the SMEFT approach. However, once the basis has been set, the result can be considered model-independent in that basis, as one only then assumes no light degrees of freedom below the unitarity-violating scale. In our approach, we eliminate derivative operators in favor of non-derivative ones to remain maximally conservative.

To proceed, it is more convenient to work in a gauge where the Goldstone-bosons are manifest so that we can use the equivalence theorem. We can do this first by writing the Higgs doublet as
\begin{equation}\label{eq:HiggsDoublet}
H = \frac{1}{\sqrt{2}} 
\begin{pmatrix}
G_{1} + i G_{2}\\
v + h + i G_{3}
\end{pmatrix},
\end{equation}
with $h$ being the physical Higgs field, and then defining the field
\begin{equation}\label{eq:Xfield}
X \equiv \sqrt{2H^{\dagger}H}-v = h+\frac{\vec{G}^{2}}{2(v+h)} - \frac{\vec{G}^{4}}{8(v+h)^{3}}+ O\Bigg( \frac{\vec{G}^{6}}{(v+h)^{5}}\Bigg),
\end{equation}
where we have introduced $\vec{G} = (G_{1},G_{2},G_{3})$\footnote{When calculating the amplitudes in this paper, we rewrite the Goldstone bosons in terms of the longitudinal modes of the $W$ and $Z$ using $W_{L}^{\pm} = \frac{1}{\sqrt{2}}(G_{1}\mp i G_{2})$ and $Z_{L} = G_{3}$.}. Because $X = h$ in the unitary gauge, we can generalize eq. (\ref{eq:effLag1}) to any general gauge by substituting $h \rightarrow X$. 

Although eq. (\ref{eq:effLag1}) appears similar to HEFT \cite{Grinstein:2007iv}, we should nonetheless keep in mind that we are implicitly restoring the Higgs doublet via the above substitution. In fact, when eq. (\ref{eq:effLag1}) is generalized to a general gauge using the aforementioned substitution, then one can show that it becomes isomorphic to SMEFT. More explicitly, the deviations $\delta_{i}$ and Wilson coefficients $c_{i}$ are assumed to receive contributions from multiple higher-order operators in the SMEFT expansion, such that one can match the generalized Lagrangian to the SMEFT expansion truncated at a certain order. To be more concrete, if we truncate SMEFT at a certain order, like dim-6 for example, and expand the operators explicitly; then one can map its parameters (i.e. the cutoff scale and Wilson coefficients in SMEFT) to the parameters in our approach (i.e., $\delta_{i}$ and $c_{i}$). The correspondence is one-to-one for a certain SMEFT truncation, but might change by including higher-order SMEFT operators. We show the matching explicitly in the $h\gamma\gamma$ and $h\gamma Z$ in Subsections \ref{Sec:hggSMEFT} and \ref{Sec:hgZSMEFT} respectively.

Operators in eq. (\ref{eq:effLag1}) give rise to amplitudes that grow with energy, which eventually violate unitarity at some high energy scale, signaling the onset of new physics. To obtain the scale of new physics $E_{\text{max}}$, we demand that the amplitudes respect unitarity up to $E_{\text{max}}$, and only assume that there are no new light degrees of freedom up to that scale. As mentioned above, we call processes that depend on one parameter only model-independent, as they do not depend on any assumptions regarding the possible UV completion. Other processes are not truly model-independent, because the scale of new physics can vary based on the assumed relations among the various parameters, which does depend on the UV completion.

As clarified in \cite{Chang:2019vez}, loops contain either heavy particles or SM particles. In the former case, the heavy particles can be integrated out to give local terms that appear in the expansion in eq. (\ref{eq:effLag1}). On the other hand, SM loops give perturbatively small corrections up to the unitarity-violating scale, where they become comparable to the tree-level contributions, and thus give $O(1)$ corrections.\footnote{In \cite{Abu-Ajamieh:2021vnh}, loops from eq. (\ref{eq:effLag1}) were studied in detail to investigate the possibility of canceling the quadratic divergences in the Higgs mass corrections. There, it was found that the quadratic divergences in the Higgs mass could be canceled if the scale of new physics is $\lesssim 19$ TeV.}

Appendix \ref{app:A1} illustrates the technical details of how the states are defined and normalized, in addition to the derivation of the unitarity condition. We present a sample calculation of the unitarity-violating scale in Appendix \ref{app:A4} to clarify the procedure. In the remainder of this section, we highlight the main results found in \cite{Chang:2019vez, Abu-Ajamieh:2020yqi}. There, the scale of new physics was investigated in the couplings $h^{3}$, $h^{4}$, $hVV$, $h^{2}VV$, $h\bar{t}t$ and $h^{2}\bar{t}t$; with $V$ referring to the longitudinal $W$ or $Z$.	

For $h^{3}$ and $h^{4}$, new physics arises from the first two terms in the BSM Lagrangian in eq. (\ref{eq:effLag1}). After restoring the Goldstone bosons, we find that the strongest model-independent bound comes from $W_{L}^{+}W_{L}^{-}W_{L}^{+} \rightarrow W_{L}^{+}W_{L}^{-}W_{L}^{+}$ and is given by
\begin{equation}\label{eq:H_Strongest1}
E_{\text{max}} \simeq \frac{14 \hspace{1mm}\text{TeV}}{|\delta_{3}|^{1/2}},
\end{equation}
and there are no model-independent bounds that only depend on the deviation in the Higgs quartic coupling $\delta_{4}$, as the operators that modify $\delta_{4}$ also modify $\delta_{3}$. Of these operators, the strongest bounds come from $W_{L}^{+}W_{L}^{-}W_{L}^{+}W_{L}^{-} \rightarrow W_{L}^{+}W_{L}^{-}W_{L}^{+}W_{L}^{-}$ and gives
\begin{equation}\label{eq:H_Strongest3}
E_{\text{max}} \simeq \frac{8.7 \hspace{1mm} \text{TeV}}{|\delta_{4} -6\delta_{3}|^{1/4}} .
\end{equation}

Notice that in eq. (\ref{eq:H_Strongest3}), the scale of physics can be varied depending on how close $\delta_{4}$ is tuned to $6\delta_{3}$, with the scale of new physics blowing up when $\delta_{4} = 6\delta_{3}$. This is simply the SM limit, which corresponds to the vanishing of the leading-order operators. In this case, the NLO operators should be included to determine the scale of new physics.

Bounds from the coupling $hVV$ depend on whether custodial symmetry is assumed or not. However, as the experimental constraints on the custodial symmetry $T$ parameter are stringent ($\delta T \lesssim 10^{-3}$), we assume that custodial symmetry is preserved, in which case $\delta_{W1} = \delta_{Z1} \equiv \delta_{V1}$. In this limit, the strongest bound arises from $Z_{L}Z_{L} \rightarrow W_{L}^{+}W_{L}^{-}$ and is given by
\begin{equation}\label{eq:WZ_Strongest_custodial}
E_{\text{max}} \simeq \frac{1.1 \hspace{1mm} \text{TeV}}{|\delta_{V1}|^{1/2}}.
\end{equation}

Bounds from the coupling $h\bar{t}t$ arise from processes that are sensitive to both $\delta_{t1}$ and $\delta_{V1}$, i.e. they are not model-independent. Nonetheless, they can be phenomenologically important, as they can be probed in the LHC or the HL-LHC, and both contribute to the di-Higgs production. The strongest bounds that are only sensitive to $\delta_{t1}$ and $\delta_{V1}$ are

\begin{equation}\label{eq:htt_strongest}
\begin{aligned}
t_{R} \bar{t}_{R} \rightarrow W_{L}^{+}W_{L}^{-} & : \hspace{2mm} E_{\text{max}} \simeq \frac{5.1 \hspace{1mm} \text{TeV}}{|\delta_{t1}+\delta_{V1}|}, \\
t_{R} \bar{b}_{R} \rightarrow W_{L}^{+}h & : \hspace{2mm} E_{\text{max}} \simeq \frac{3.6 \hspace{1mm} \text{TeV}}{|\delta_{t1}-\delta_{V1}|}, \\
t_{R} \bar{b}_{R} \rightarrow W_{L}^{+}W_{L}^{-}W_{L}^{+} & : \hspace{2mm} E_{\text{max}} \simeq \frac{3.3 \hspace{1mm} \text{TeV}}{\sqrt{|\delta_{t1}-\frac{1}{3}\delta_{V1}|}} . 
\end{aligned}
\end{equation}

Similar to the coupling $h\bar{t}t$, bounds from the coupling $h^{2}VV$ are not model-independent as they are sensitive to $\delta_{V1}$ as well. The strongest bounds that depend on both $\delta_{V1}$ and $\delta_{V2}$ arise from
\begin{equation}\label{eq:hhVV_strongest}
\begin{aligned}
W_{L}^{+}W_{L}^{-} \rightarrow hh & :  \hspace{2mm} E_{\text{max}} \simeq \frac{1.5 \hspace{1mm} \text{TeV}}{|\delta_{V2}-2\delta_{V1}|^{1/2}}, \\
Z_{L}Z_{L}\rightarrow hW_{L}^{+}W_{L}^{-} & :  \hspace{2mm} E_{\text{max}} \simeq \frac{1.9 \hspace{1mm} \text{TeV}}{|\delta_{V2}-4\delta_{V1}|^{1/3}}, \\
W_{L}^{+}W_{L}^{+}Z_{L} \rightarrow W_{L}^{+}W_{L}^{+}Z_{L} & : \hspace{2mm} E_{\text{max}} \simeq \frac{2.6 \hspace{1mm} \text{TeV}}{|\delta_{V2}-4\delta_{V1}|^{1/4}}. 
\end{aligned}
\end{equation}
and finally, the strongest bounds associated with the coupling $h^{2}\bar{t}t$ arise from
\begin{equation}\label{eq:hhtt_strongest}
\begin{aligned}
t_{R}\bar{t}_{R} \rightarrow hh & :  \hspace{2mm} E_{\text{max}} \simeq \frac{7.2 \hspace{1mm} \text{TeV}}{|c_{t2}|}, \\
t_{R}\bar{t}_{R} \rightarrow W_{L}^{+}W_{L}^{-	}h & : \hspace{2mm} E_{\text{max}} \simeq \frac{4.7 \hspace{1mm} \text{TeV}}{|c_{t2}-2\delta_{t1}+\frac{1}{3}\delta_{V2}|^{1/2}}, \\
t_{R}\bar{b}_{R} \rightarrow W_{L}^{+}hh & : \hspace{2mm} E_{\text{max}} \simeq \frac{4.7 \hspace{1mm} \text{TeV}}{|c_{t2}-2\delta_{t1}-\frac{2}{3}\delta_{V2}|^{1/2}}, \\
t_{R}\bar{b}_{R}W_{L}^{-} \rightarrow hW_{L}^{+}W_{L}^{-} & : \hspace{2mm} E_{\text{max}} \simeq \frac{3.9 \hspace{1mm} \text{TeV}}{|c_{t2}-3\delta_{t1}+\frac{1}{2}\delta_{V2}|^{1/3}}, \\
t_{R}\bar{b}_{R}W_{L}^{-} \rightarrow W_{L}^{+}W_{L}^{-}W_{L}^{+}W_{L}^{-} & : \hspace{2mm} E_{\text{max}} \simeq \frac{4.2 \hspace{1mm} \text{TeV}}{|c_{t2}-3\delta_{t1}+\frac{1}{3}\delta_{V2}|^{1/4}}. 
\end{aligned}
\end{equation}

Given the current constraint and HL-LHC projections on $\delta_{3}$,  $\delta_{4}$,  $\delta_{V1}$, $\delta_{V2}$, $\delta_{t1}$ and $c_{t2}$ \cite{Cepeda:2019klc, ATLAS:2019nkf, ATLAS:2018jlh, Bishara:2016kjn, ATLAS:2020jgy, Azatov:2015oxa}; it was found in \cite{Chang:2019vez, Abu-Ajamieh:2020yqi} that the current limits on the coupling $hVV$ allow for new physics below $\sim 3$ TeV, whereas the limits on the $h\bar{t}t$ coupling allow for new physics below $\sim 8$ TeV. On the other hand, the deviation in Higgs trilinear coupling $\delta_{3}$ allows for new physics below $\sim 4$ TeV. The couplings $h^{2}VV$ and $h^{2}\bar{t}t$ are particularly important for di-Higgs searches produced via gluon fusion and vector boson fusion, and their constraints allow for new physics as low as $\sim 2$ TeV. All these scales should be accessible in the HL-LHC.

\section{New Physics from the $\gamma\gamma$ Sector}
\label{Sec:hgg}
In this section, we extend the treatment in \cite{Chang:2019vez, Abu-Ajamieh:2020yqi} to the Higgs interaction with a pair of photons. As the Higgs couples to $\gamma\gamma$ at loop-level, we would expect the scale of new physics to be larger than the tree-level ones presented in Section \ref{Sec:Review}. Nonetheless, measuring the coupling $h\gamma\gamma$ to probe new physics is of particular interest, as the Higgs interaction with photons has a low background and a unique signature.

\subsection{Model-Independent Bound on the Scale of New Physics from the Coupling $h\gamma\gamma$}\label{Sec:hggBound1}
\begin{figure}[!t]
\centerline{\begin{minipage}{0.5\textwidth}
\centerline{\includegraphics[width=300pt]{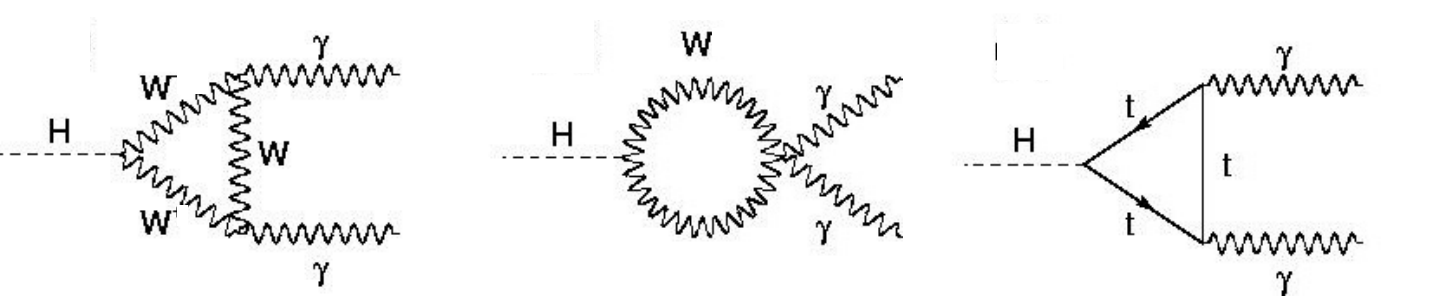}}
\caption{\small The SM coupling of the Higgs to $\gamma\gamma$.}
\label{fig2}
\end{minipage}}
\end{figure}

In the SM, the Higgs couples to a pair of photons through a fermion loop or a $W$ loop as shown in Figure \ref{fig2}. The contribution of the top quark dominates over all other fermions due to its larger Yukawa coupling, whereas the contribution from the $W$ loops dominates over the fermionic one. One can integrate out the fermion and $W$ loops to write an effective tree-level operator for $h\gamma\gamma$. Using the convention in \cite{Carmi:2012in}, one can write the effective Lagrangian in the unitary gauge as
\begin{equation}\label{eq:Hgaga_Lag1}
\mathcal{L}_{h\gamma\gamma} = c_{\gamma 1} \frac{\alpha}{\pi v} h A_{\mu\nu}A^{\mu\nu},
\end{equation}
where $A_{\mu\nu} = \partial_{\mu}A_{\nu} - \partial_{\nu}A_{\mu}$, and in the SM, $c^{\text{SM}}_{\gamma 1} \simeq 0.81$ when only the top quark is retained. Similar to eq. (\ref{eq:delta}), it is more convenient to express new physics in terms of the deviation in the $h\gamma\gamma$ coupling. Thus, we define
\begin{equation}\label{eq:dev1}
\delta_{\gamma 1} \equiv \frac{c_{\gamma 1}-c^{\text{SM}}_{\gamma 1}}{c^{\text{SM}}_{\gamma 1}}.
\end{equation}

Moving from the unitary gauge by restoring the Goldstones using eq. (\ref{eq:Xfield}), and expressing $h\gamma\gamma$ in terms of $\delta_{\gamma 1}$, we write eq. (\ref{eq:Hgaga_Lag1}) as
\begin{equation}\label{eq:Hgaga_Lag2}
\mathcal{L}_{h\gamma\gamma} = \Big(\frac{\alpha c^{\text{SM} }_{\gamma 1} \delta_{\gamma 1}}{2\pi v^{2}}\Big) A_{\mu\nu}A^{\mu\nu} \Big[Z_{L}^{2} + 2 W_{L}^{+}W_{L}^{-} \Big]+ \dots,
\end{equation}
where the ellipsis indicates higher-order operators that depend on other deviations and/ or Wilson coefficients. In this paper, we limit ourselves to the contact interactions and neglect processes with propagators.\footnote{For a complete discussion of processes with propagators and the possible IR enhancement, see \cite{Abu-Ajamieh:2020yqi}.} The strongest bound comes from $\gamma_{\pm} \gamma_{\pm} \rightarrow W_{L}^{+}W_{L}^{-}$ and reads
\begin{equation}\label{eq:delta1_strongest}
E_{\text{max}} \simeq \frac{23.9 \hspace{1mm} \text{TeV}}{|\delta_{\gamma 1}|^{1/2}}.
\end{equation}

\begin{figure}[!t]
\centerline{\begin{minipage}{0.8\textwidth}
\centerline{\includegraphics[width=300pt]{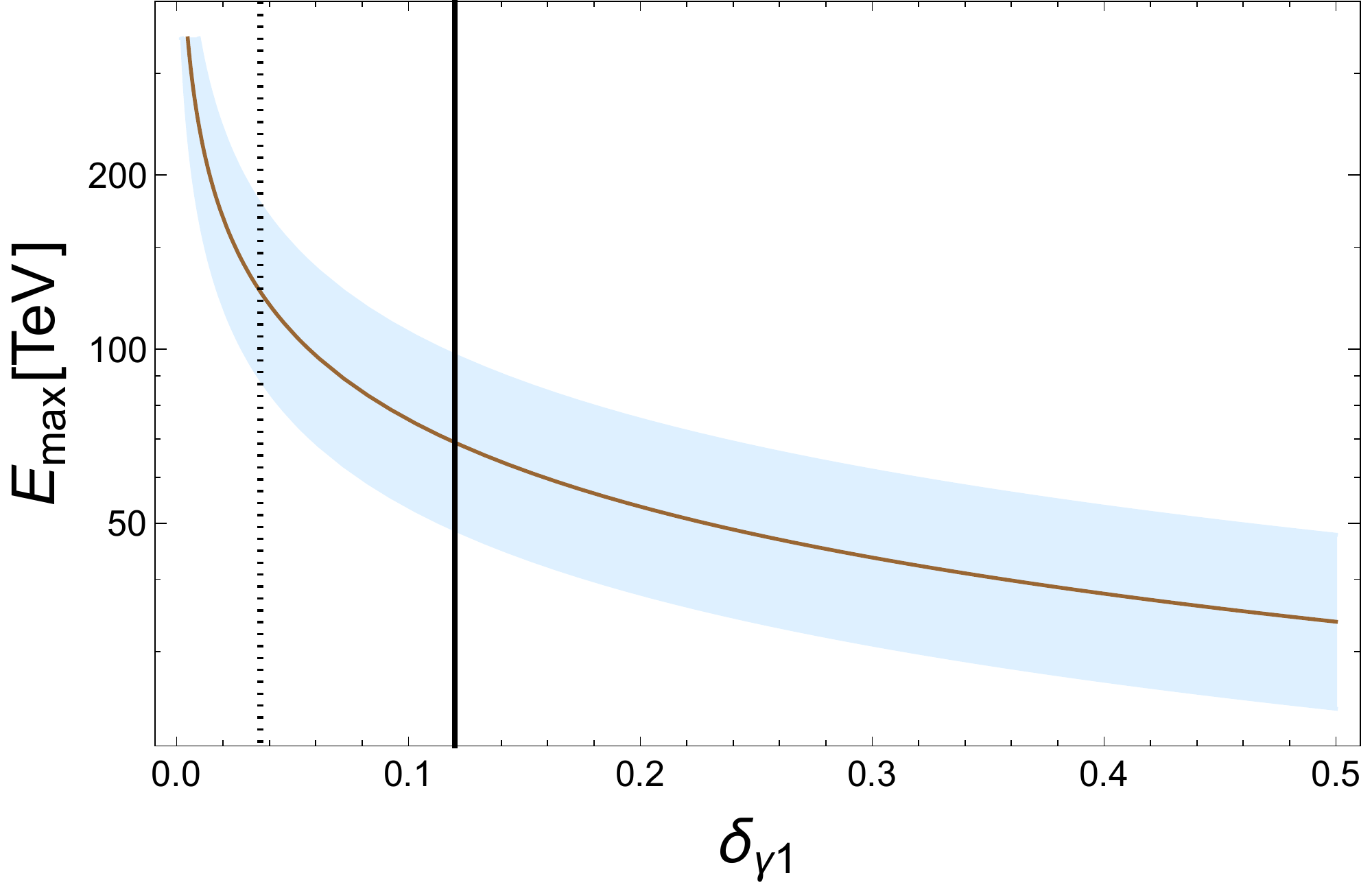}}
\caption{\small 
The unitarity-violating bound as a function of $\delta_{\gamma 1}$. The model-independent bound is obtained from eq. (\ref{eq:delta1_strongest}). The band around the model-independent bound results from varying the unitarity bound to $\frac{1}{2} \leq \hat{\mathscr{M}} \leq 2$. The solid black line represents the $95\%$ C.L. limits on $\delta_{\gamma 1}$ obtained from ATLAS \cite{ATLAS:2019nkf}, whereas the dotted one represents the $95\%$ C.L. HL-LHC projections obtained from \cite{Cepeda:2019klc}.}
\label{fig3}
\end{minipage}}
\end{figure}

ATLAS \cite{ATLAS:2019nkf} places the $95\%$ limits on $\delta_{\gamma 1}$ at $\pm 12\%$. This places the scale of new physics at $\sim 69$ TeV at best, which is well beyond what can be probed in the LHC, although it might be marginally within the reach of the $100$ TeV collider. However, $\delta_{\gamma 1}$ can be investigated indirectly through precision measurements at the HL-LHC, which would help set limits on the scale of new physics arising from the coupling $h\gamma\gamma$. For instance, \cite{Cepeda:2019klc} puts the HL-LHC $95\%$ projections for $\delta_{\gamma 1}$ at $\sim \pm 3.6\%$, which would push the scale of new physics above $\sim 125$ TeV. We plot the unitarity bound as a function of $\delta_{\gamma 1}$ in Figure \ref{fig3}, together with the LHC limits and HL-LHC projections.

\subsection{Model-Independent Bounds on the Scale of New Physics from the Couplings $h^{n}\gamma\gamma$}\label{Sec:hggBoundn}
\begin{figure}[!t]
\centerline{\begin{minipage}{0.6\textwidth}
\centerline{\includegraphics[width=250pt]{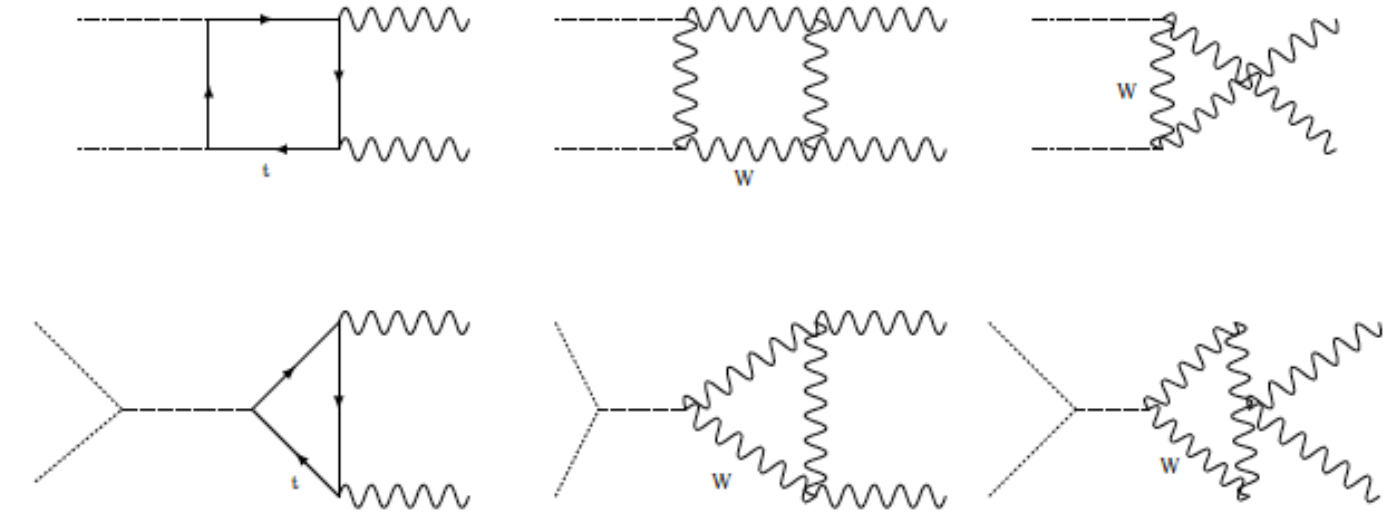}}
\caption{\small The SM coupling of two Higgses to $\gamma\gamma$.}
\label{fig4}
\end{minipage}}
\end{figure}

We can easily generalize eq. (\ref{eq:Hgaga_Lag1}) to include the interactions of any number of Higgses with two photons by writing the effective Lagrangian as

\begin{equation}\label{eq:nHgaga_Lag1}
\mathcal{L}_{h^{n}\gamma\gamma} = A_{\mu\nu}A^{\mu\nu} \sum_{n=1}^{\infty}\frac{c_{\gamma n}}{n!}\frac{\alpha}{\pi}\Big( \frac{h}{v} \Big)^{n},
\end{equation}
where $c_{\gamma n}$ are the Wilson coefficients that parametrize the effective couplings of $n$-Higgses with $\gamma\gamma$. Notice that $c_{\gamma n}$ are non-vanishing in the SM, for instance, in the SM, $c_{\gamma 2}$ arises from integrating out the loops (and propagators) of the diagrams shown in Figure \ref{fig4}. However, unlike $c^{\text{SM}}_{\gamma 1}$, $c^{\text{SM}}_{\gamma n}$ for $n \geq 2$ have never been calculated to the best of our knowledge. Thus, it is more convenient to define the deviations $\delta_{\gamma n}$ in terms of $c^{\text{SM}}_{\gamma 1}$ instead
\begin{equation}\label{eq:delta_ga_n}
\delta_{\gamma n} \equiv \frac{c_{\gamma n}-c^{\text{SM}}_{\gamma n}}{c^{\text{SM}}_{\gamma 1}}, \hspace{5mm} \text{for} \hspace{2mm} n \geq 2,
\end{equation}
and treat $\delta_{\gamma n}$ as free parameters. Although defining $\delta_{\gamma n}$ this way does not help us quantify precisely the deviation from the SM, it does allow us nonetheless to express the SM limit more transparently as $\delta_{\gamma n} \rightarrow 0$. Restoring the Goldstone bosons, eq. (\ref{eq:nHgaga_Lag1}) becomes
\begin{dmath}\label{eq:nHgaga_Lag2}
\mathcal{L}_{h^{n}\gamma\gamma} = \frac{\alpha c^{\text{SM} }_{\gamma 1}}{2\pi} A_{\mu\nu}A^{\mu\nu} \Big\{ \Big(\frac{\delta_{\gamma 1}}{v^{2}} \Big) \Big[Z_{L}^{2} + 2 W_{L}^{+}W_{L}^{-} \Big] + \Big(\frac{\delta_{\gamma 2}}{v^{2}} \Big)h^{2} + 
 \Big( \frac{\delta_{\gamma 2}-\delta_{\gamma 1}}{v^{3}}\Big)h \Big[Z_{L}^{2} + 2 W_{L}^{+}W_{L}^{-} \Big]  + \Big(\frac{\delta_{\gamma 3}}{3v^{3}} \Big)h^{3} + 
 \Big( \frac{\delta_{\gamma 2}- \delta_{\gamma 1}}{4v^{4}}\Big) \Big[Z_{L}^{2} + 2 W_{L}^{+}W_{L}^{-} \Big]^{2}+ 
 \Big( \frac{\delta_{\gamma 1}-\delta_{\gamma 2}+\frac{1}{2}\delta_{\gamma3}}{v^{4}}\Big) h^{2} \Big[  Z_{L}^{2} + 2 W_{L}^{+}W_{L}^{-}\Big] + \Big(\frac{\delta_{\gamma 4}}{12v^{4}} \Big)h^{4} + \dots 
\Big\}.
\end{dmath}

Notice here that eq. (\ref{eq:Hgaga_Lag2}) can be obtained from eq. (\ref{eq:nHgaga_Lag2}) by setting $\delta_{\gamma n} \rightarrow 0$ for $n > 1$. Up to the dim-8 operators shown above, the strongest model-independent bounds are given by the following processes
\begin{dmath}\label{eq:nHgaga_Strongest}
\begin{aligned}
\gamma_{\pm}\gamma_{\pm} \rightarrow hh, \gamma_{\pm}h \rightarrow \gamma_{\mp}h & :  \hspace{2mm} E_{\text{max}} \simeq \frac{28.4 \hspace{1mm}\text{TeV}}{|\delta_{\gamma 2}|^{1/2}}, \\
\gamma_{\pm}\gamma_{\pm} \rightarrow hhh & :  \hspace{2mm} E_{\text{max}} \simeq \frac{18.3 \hspace{1mm}\text{TeV}}{|\delta_{\gamma 3}|^{1/3}}, \\
\gamma_{\pm}hh \rightarrow \gamma_{\mp}hh & : \hspace{2mm} E_{\text{max}} \simeq \frac{16.2 \hspace{1mm}\text{TeV}}{|\delta_{\gamma 4}|^{1/4}}. \\
\end{aligned}
\end{dmath}

Given that $\delta_{\gamma n}$ are essentially free parameters, eq. (\ref{eq:nHgaga_Strongest}) seems to suggest that the scale of new physics could be significantly low and perhaps even within the reach of the LHC. However, this would require $\delta_{\gamma n}$ to be unnaturally large. For instance, bringing $E_{\text{max}}$ to be $\lesssim 10$ TeV would require $\delta_{\gamma 2,3,4} \gtrsim 7$, which is extremely unlikely despite not being ruled out by experiment. In spite of this, the fact that the limits on the LO deviation $\delta_{\gamma 1}$ only allow for $E_{\text{max}} \gtrsim 69$ TeV makes investigating the NLO deviations more promising to search for new physics, as $E_{\text{max}}$ could be within the reach of the $100$ TeV collider for moderate values of $\delta_{\gamma n}$. For instance, for $\delta_{\gamma 3,4} \sim 0.2$, the scale of new physics could be $\lesssim 31$ TeV, which is significantly lower than the scale from $\delta_{\gamma 1}$.

Notice that the processes in eq. (\ref{eq:nHgaga_Strongest}) are not the only model-independent ones that can be obtained from the Lagrangian in eq. (\ref{eq:nHgaga_Lag2}). In fact, it is not hard to see that all operators that involve $\gamma\gamma$ with any number of pure Higgses are indeed model-independent, as they only depend on a single deviation $\delta_{\gamma n}$. More concretely, we consider the following operators
\begin{equation}\label{eq:2n+2_model_indep}
\delta \mathcal{L}^{\gamma\gamma}_{\text{2n+2}} = \Big( \frac{c_{\gamma 1}^{\text{SM}}\delta_{\gamma 2n}\alpha}{(2n)!\pi}\Big) \Big( \frac{h}{v}\Big)^{2n} A_{\mu\nu}A^{\mu\nu}.
\end{equation}

To investigate the unitarity-violating scale, we first notice that the strongest bounds are obtained from symmetric processes, a fact that stems purely from combinatorics as shown in detail in \cite{Chang:2019vez}. Thus, we only need to consider processes with an equal number of particles in the initial and final states. We find the following bounds
\begin{dmath}\label{eq:2n_bounds}
\begin{aligned}
\gamma_{\pm}\gamma_{\pm} h^{n-1}\rightarrow h^{n+1} & :  \hspace{2mm} E_{\text{max}} = 4\pi v \Bigg( \frac{(n+1)!n!\sqrt{(n+1)!(n-1)!}}{4\sqrt{2} \alpha c^{\text{SM}}_{\gamma 1}|\delta_{\gamma 2n}|}\Bigg)^{\frac{1}{2n}}, \\[2mm]
\gamma_{\pm}h^{n}\rightarrow \gamma_{\mp}  h^{n} & : \hspace{2mm}  E_{\text{max}} = 4\pi v \Bigg( \frac{(n+1)!(n+1)!(n-1)!}{8\alpha c^{\text{SM}}_{\gamma 1} |\delta_{\gamma 2n}|}\Bigg)^{\frac{1}{2n}}. \\
\end{aligned}
\end{dmath}

\begin{figure}[!t]
\centerline{\begin{minipage}{0.8\textwidth}
\centerline{\includegraphics[width=250pt]{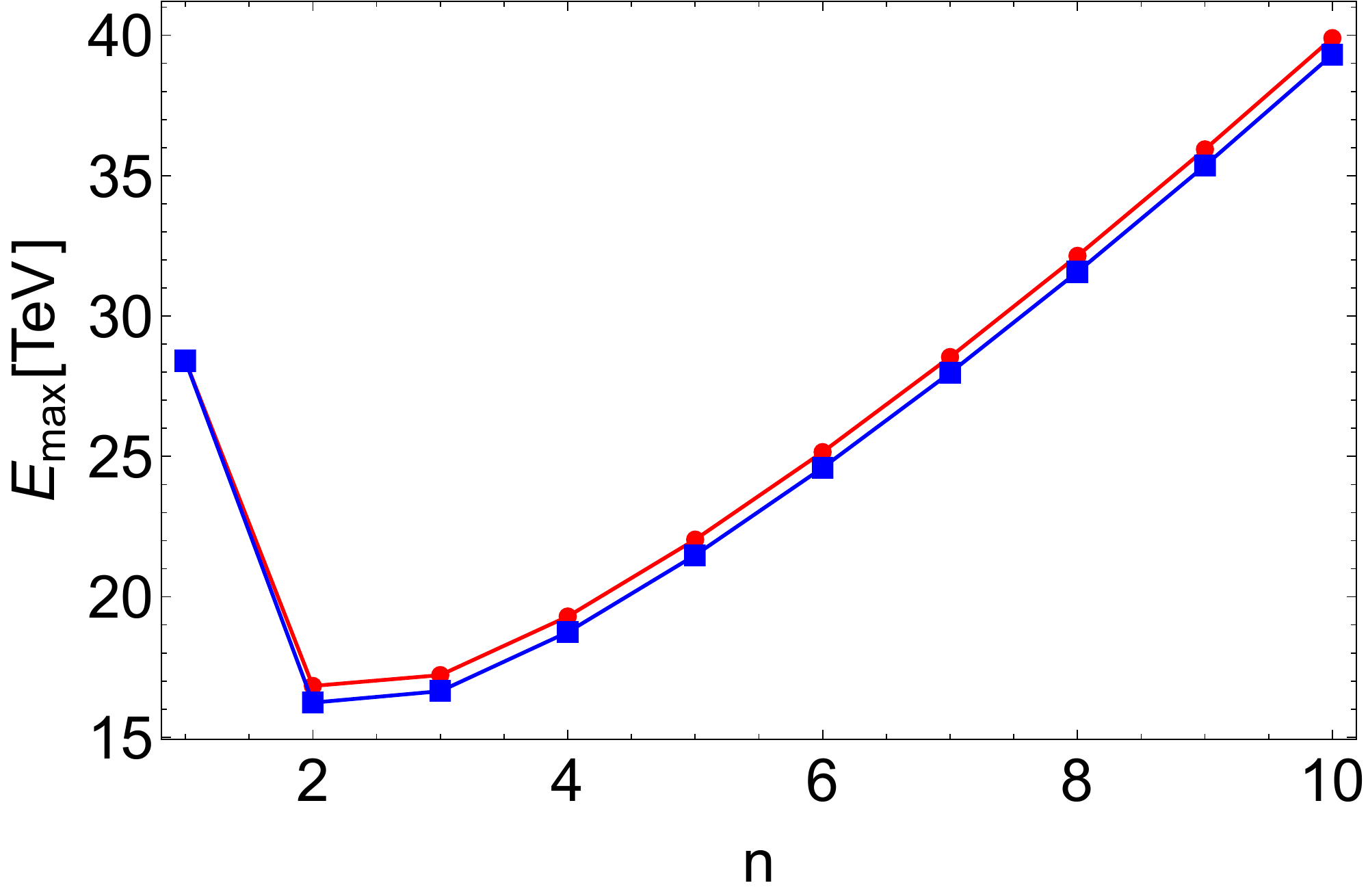}}
\caption{\small 
The unitarity-violating scale as a function of $n$ from $\gamma_{\pm} h^{n} \rightarrow \gamma_{\mp} h^{n}$ (blue), and $\gamma_{\pm} \gamma_{\pm} h^{n-1} \rightarrow h^{n+1}$ (red). The former processes are dominant for all $n$. Here $\delta_{\gamma 2n}$ is set to $1$.}
\label{fig5}
\end{minipage}}
\end{figure}

Figure \ref{fig5} shows a comparison between the two processes in eq. (\ref{eq:2n_bounds}), where we can see that $\gamma_{\pm} h^{n} \rightarrow \gamma_{\mp} h^{n}$ always dominates for all $n$. The plot also shows an interesting behavior; the unitarity bound decreases with $n$ down to a minimum at a certain $n$, before it starts increasing again. This seems to suggest that the strongest bound does not continue to decrease indefinitely as the number of scattering particles increases, as one would anticipate naively from combinatorics considerations. This suggests that the lowest unitarity violating scale is not far from the ones given in eq. (\ref{eq:nHgaga_Strongest}). However, there is a caveat to this conclusion: In Figure \ref{fig5} we set $\delta_{\gamma 2n} =1$, and different values might change this behavior. However, we checked that this behavior is maintained as long as $\delta_{\gamma 2n}$ becomes smaller with $n$, which is quite a conservative assumption. Furthermore, we also checked that the minimum is not too sensitive to specific values of $\delta_{\gamma 2n}$ and remains around $n \sim 2-3$. Thus, we conclude that the strongest unitarity-violating scale should not be too far from the ones given in eq. (\ref{eq:nHgaga_Strongest}).

\subsection{Other Bounds}\label{Sec:hggMixedProcesses}
Although the processes given in eqs. (\ref{eq:delta1_strongest}), (\ref{eq:nHgaga_Strongest}) and (\ref{eq:2n_bounds}) are the only model-independent ones, other processes that are not model-independent are still worthwhile investigating. The reason for this is that for a certain choice of the NLO deviations, the scale of new physics could be lower than the model-independent ones, and could potentially be probed in the HL-LHC. In addition, these processes can help constrain higher-order deviations that are usually harder to measure directly. Here, we limit ourselves to the unitarity-violating processes that depend only on $\delta_{\gamma 1}$ and $\delta_{\gamma 2}$. Using the general Lagrangian in eq. (\ref{eq:nHgaga_Lag2}), the strongest bounds are 
\begin{dmath}\label{eq:NLObounds}
\begin{aligned}
\gamma_{\pm}\gamma_{\pm} \rightarrow W_{L}^{+}W_{L}^{-} h & : \hspace{2mm} E_{\text{max}} \simeq \frac{13.6 \hspace{1mm}\text{TeV}}{|\delta_{\gamma 2} - \delta_{\gamma 1}|^{1/3}}, \\
\gamma_{\pm} W_{L}^{+}W_{L}^{-} \rightarrow \gamma_{\mp} W_{L}^{+}W_{L}^{-}  & :  \hspace{2mm} E_{\text{max}} \simeq \frac{11.5 \hspace{1mm}\text{TeV}}{|\delta_{\gamma 2} - \delta_{\gamma 1}|^{1/4}}. \\
\end{aligned}
\end{dmath}

\begin{figure}[!t]
\centerline{\begin{minipage}{0.8\textwidth}
\centerline{\includegraphics[width=300pt]{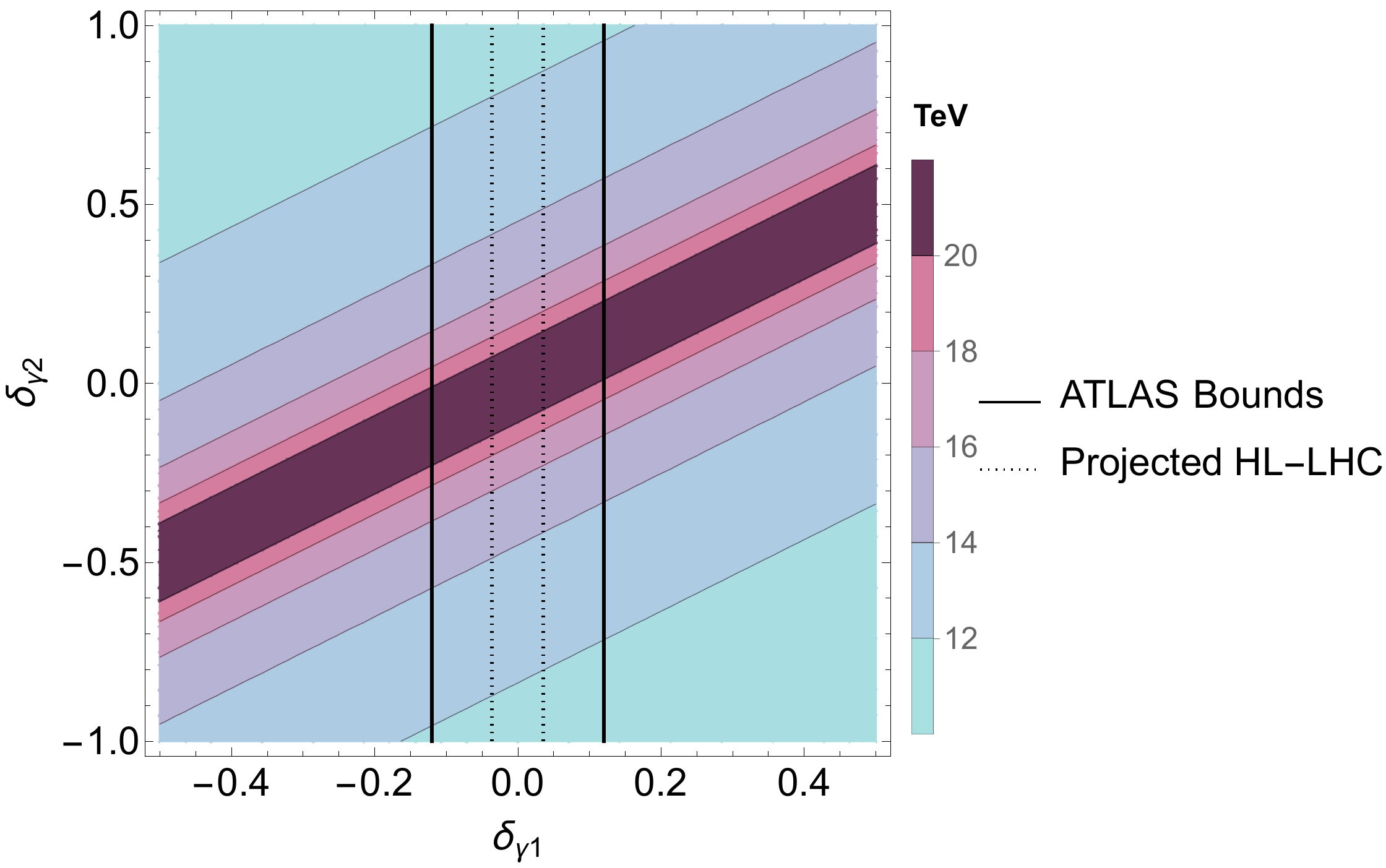}}
\caption{\small 
The unitarity-violating scale that depends on both $\delta_{\gamma 1}$ and $\delta_{\gamma 2}$. The solid black line represents the $95\%$ C.L. limits on $\delta_{\gamma 1}$ from ATLAS \cite{ATLAS:2019nkf}, whereas the dotted line represents the $95\%$ HL-LHC projections \cite{Cepeda:2019klc}. Here we set an upper limit on $|\delta_{\gamma 2}| \leq 1$.}
\label{fig6}
\end{minipage}}
\end{figure}

Figure \ref{fig6} shows the strongest unitarity-violating scale from eq. (\ref{eq:NLObounds}), with the ATLAS  limits \cite{ATLAS:2019nkf} and HL-LHC projections \cite{Cepeda:2019klc} superimposed. The positive diagonal corresponds to $\delta_{\gamma 2} =\delta_{\gamma 1}$ where the unitarity-violating scale blows up as can be seen from eq. (\ref{eq:NLObounds}). This simply means that the corresponding operators in (\ref{eq:nHgaga_Lag2}) vanish and that the NLO operators need to be included. The plot also shows that the scale of new physics could be smaller than the model-independent ones. For example, if we make the conservative assumption that $|\delta_{\gamma 2}| \leq 1$, then the scale of new physics could be as low as $\sim 12$ TeV. Although this is still beyond the reach of the HL-LHC, it is more accessible in future colliders such as the $100$ TeV collider. In addition, one can use the limits on $\delta_{\gamma 1}$ to set bound $\delta_{\gamma 2}$ using any null collider searches.
% ---------------------------------------------------------------------------------
\subsection{SMEFT Predictions from Unitarity}
\label{Sec:hggSMEFT}
As mentioned in the introduction, the SMEFT approach is predicated upon the assumption that the UV theory is of the decoupling type, which means that in the low-energy theory, the effects of new physics should be captured by adding to the SM higher-dimensional gauge-invariant operators that become more and more irrelevant the higher the scale of new physics becomes. As we mentioned before, SMEFT does not provide a way to estimate the uncertainty associated with neglecting higher-order operators. It was shown in \cite{Abu-Ajamieh:2020yqi} that unitarity arguments alone can be used to make a quantitative statement about the accuracy of SMEFT.

To do this, we should note that for the SMEFT description to be sound, then dim-6 SMEFT operators need to dominate over higher-order ones. To be quantitative, we consider a theory consisting of the SM in addition to the dim-6 SMEFT operator
\begin{equation}\label{eq:ga_dim6_SMEFT}
\mathcal{L}^{\text{dim-6}}_{\text{SMEFT}} = \frac{1}{M^{2}}\Big( H^{\dagger}H - \frac{v^{2}}{2}\Big) A_{\mu\nu}A^{\mu\nu},
\end{equation}
and demand that it dominates over all higher-order operators. From this, we find that SMEFT predicts the following relations
\begin{equation}\label{eq:ga_SMEFTpred}
\delta_{\gamma 1} = \frac{\pi v^{2}}{\alpha c^{\text{SM}}_{\gamma 1}M^{2}}, \hspace{10mm} \delta_{\gamma 2} = \delta_{\gamma 1}, \hspace{10mm} \delta_{\gamma n} = 0 \hspace{2 mm} \text{for} \hspace{2mm} n  \geq 3.
\end{equation}

\begin{figure}[!t]
\centerline{\begin{minipage}{0.8\textwidth}
\centerline{\includegraphics[width=300pt]{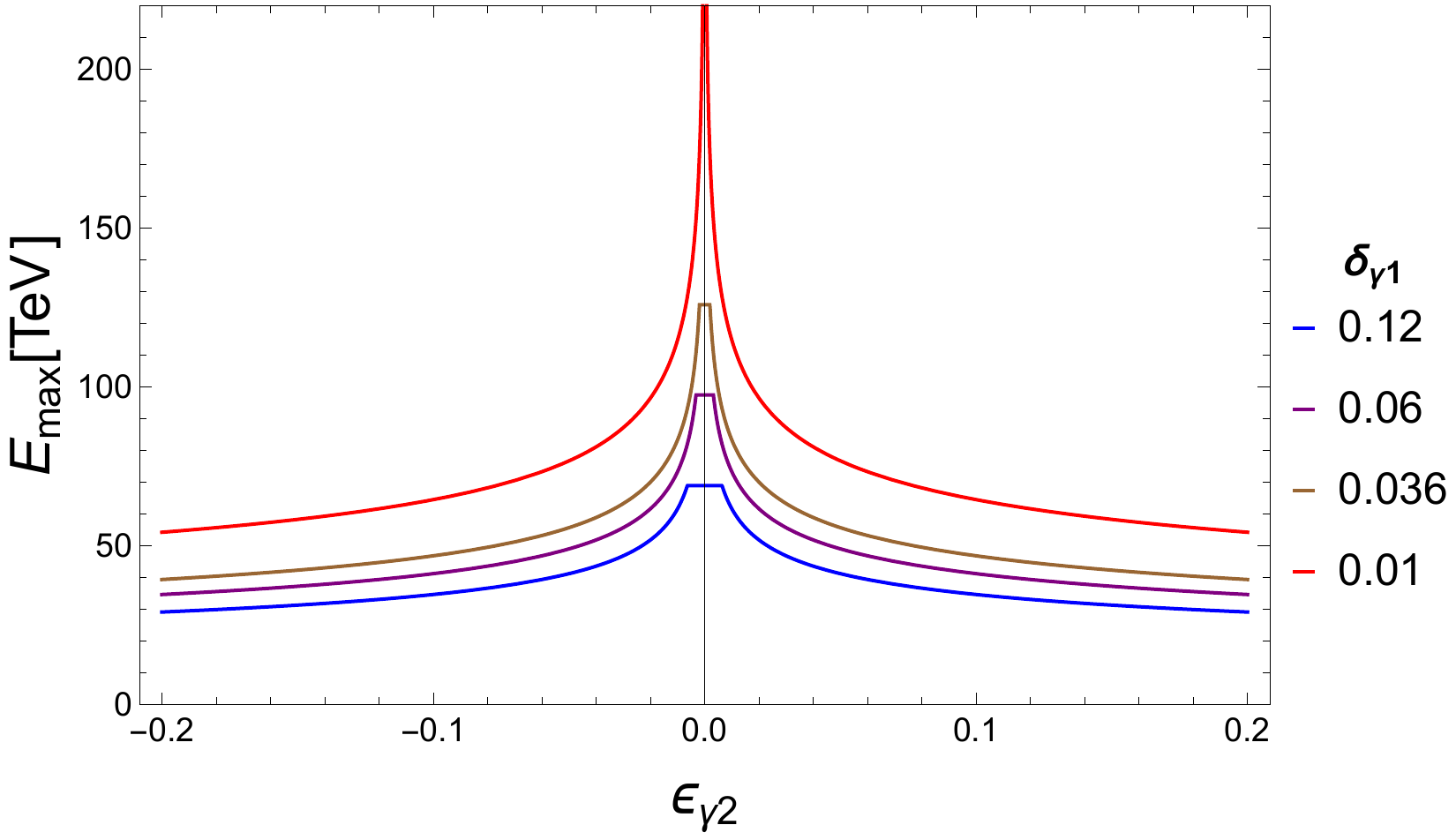}}
\caption{\small 
The unitarity-violating scale from processes that depend on both $\delta_{\gamma 1}$ and $\delta_{\gamma 2}$ as function of the fractional deviation $\epsilon_{\gamma 2}$ from the predictions of the dim-6 SMEFT operator. See eqs. (\ref{eq:ga_SMEFTpred}) and (\ref{eq:ga_epsilon}).}
\label{fig7}
\end{minipage}}
\end{figure}

To estimate the accuracy of SMEFT, we require that the scale of new physics that comes from $\delta_{\gamma 2}$ does not fall below that which comes from  $\delta_{\gamma 1}$. This requirement necessitates that $\delta_{\gamma 2}$ be tuned to be close to the SMEFT prediction in eq. (\ref{eq:ga_SMEFTpred}). This can most easily be seen by inspecting eq. (\ref{eq:NLObounds}), where we can see that the unitarity-violating scale becomes larger when $\delta_{\gamma 2}$ is tuned to be close to $\delta_{\gamma 1}$. To be more quantitative, we parametrize the deviation in $\delta_{\gamma 2}$ from the SMEFT prediction as
\begin{equation}\label{eq:ga_epsilon}
\epsilon_{\gamma 2} \equiv \frac{\delta_{\gamma 2} - \delta^{\text{SMEFT}}_{\gamma 2}}{\delta^{\text{SMEFT}}_{\gamma 2}} \ll 1,
\end{equation}
where $\delta_{\gamma 2}^{\text{SMEFT}}$ is given by eq. (\ref{eq:ga_SMEFTpred}). The process that is most sensitive to new physics is given by the second line in eq. (\ref{eq:NLObounds}). We plot the bound from this process against the deviation from the SMEFT predictions $\epsilon_{\gamma 2}$ in Figure \ref{fig7}. The plot shows that the higher the scale of new physics is, the more accurate SMEFT becomes. For example, Suppose the HL-LHC measures $|\delta_{\gamma 1}|$ to be within $3.6 \%$ as projected in \cite{Cepeda:2019klc}, then eq. (\ref{eq:delta1_strongest}) would predict a scale of new physics above $125$ TeV, which in turn would limit $|\delta_{\gamma 2}|$ to be less than $\sim 2 \times 10^{-3}$ of the SMEFT predictions as can be seen from Figure \ref{fig7}. Thus, not only do precision measurements of $\delta_{\gamma 1}$ predict the scale of new physics, but they also can help constrain $\delta_{\gamma 2}$, which is hard to measure directly.

\section{New Physics from the $\gamma Z$ Sector}
\label{Sec:hgZ}

In this section, we investigate new physics in the $\gamma Z$ sector. In the SM, the Higgs couples to $\gamma Z$ in exactly the same way as it couples to $\gamma\gamma$ (same as in Figure \ref{fig2}, but with one of the final states photons being replaced with a $Z$). Thus, we would naively expect the bounds to be somewhat similar to the ones obtained in the $\gamma\gamma$ sector. In our calculation, we only retain the top quark in the fermion loop, as usual.

\subsection{Model-Independent Bound on the Scale of New Physics from the Coupling $h\gamma Z$}\label{Sec:hgZBound}

As the coupling of the Higgs to $\gamma Z$ in the SM is very similar to its coupling to $\gamma\gamma$, we can write the effective interaction after integrating out the loops in a manner similar to eq. (\ref{eq:Hgaga_Lag1}). In the unitary gauge, we have
\begin{equation}\label{eq:HgaZ_Lag1}
\mathcal{L}_{\gamma Z} = c_{\gamma Z1}\frac{\alpha}{\pi v}h A_{\mu\nu} Z^{\mu\nu},
\end{equation} 
where in the SM, $c^{\text{SM}}_{\gamma Z1} \simeq 2.87$ when only the contribution of the top quark is retained. We define the deviation in the coupling in the usual way
\begin{equation}\label{eq:hgaZ_dev}
\delta_{\gamma Z1} \equiv \frac{c_{\gamma Z1}-c^{\text{SM}}_{\gamma Z1}}{c^{\text{SM}}_{\gamma Z1}}.	
\end{equation}

The Goldstone bosons can be restored in the Higgs field in the usual way as in eq. (\ref{eq:Xfield}). On the other hand, one also needs to restore the Goldstone bosons explicitly in the $Z$ field\footnote{In \cite{Abu-Ajamieh:2020yqi}, the explicit Goldstone dependence is restored in the $Z$ field using the following projector:
\begin{equation}\label{eq:Zpojector}
\hat{H}^{\dagger}iD^{\mu}\hat{H} = -\frac{m_{Z}}{v}Z^{\mu} -\frac{1}{v}\partial^{\mu}G_{3} + \dots,
\end{equation} 
where $\hat{H} = \frac{H}{\sqrt{H^{\dagger}H}}$.}, but before we do this, it is convenient to first separate the longitudinal and transverse modes of the $Z$
\begin{equation}\label{eq:Zexplicit}
Z^{\mu} = Z^{\mu}_{L} + Z^{\mu}_{T}.
\end{equation}

Notice that when we plug eq. (\ref{eq:Zexplicit}) in eq. (\ref{eq:HgaZ_Lag1}), we can immediately see that the longitudinal part of $Z$ does not conserve angular momentum. Thus, all operators containing $Z_{L}$ must have vanishing amplitudes, and only the transverse mode $Z_{T}$ contributes to the amplitudes. So from now on, we drop $Z_{L}$. Finally, we obtain
\begin{equation}\label{eq:HgaZ_Lag2}
\mathcal{L}_{h\gamma Z} = \Big(\frac{\alpha c^{\text{SM} }_{\gamma Z 1} \delta_{\gamma Z1}}{2\pi v^{2}}\Big) A_{\mu\nu}Z^{\mu\nu}_{T} \Big[Z_{L}^{2} + 2 W_{L}^{+}W_{L}^{-} \Big]+ \dots
\end{equation}

Due to the similarity of the $h\gamma Z$ operator to the $h \gamma \gamma$ operator, we can immediately read-off the amplitudes after making the proper replacements of the couplings and deviations, and after accounting for the combinatorics factors and normalization constants that are slightly different due to the gauge bosons in eq. (\ref{eq:HgaZ_Lag2}) no longer being identical particles. The strongest bound from eq. (\ref{eq:HgaZ_Lag2}) comes from $\gamma_{\pm} Z_{T \pm} \rightarrow W_{L}^{+}W_{L}^{-}$ and gives the bound
\begin{equation}\label{eq:deltaZ1_strongest}
E_{\text{max}} \simeq \frac{15.1 \hspace{1mm} \text{TeV}}{|\delta_{\gamma Z1}|^{1/2}}.
\end{equation}

ATLAS \cite{ATLAS:2015egz} puts the $95\%$ C.L. on $\delta_{\gamma Z1}$ at $\pm 2.3$, which allows for a scale of new physics that could lie below $\sim 10$ TeV. However, we should keep in mind that this is probably an artifact of the weak constraints on $\delta_{\gamma Z1}$, which is unlikely to be much larger than $\delta_{\gamma 1}$. For example, if $\delta_{\gamma Z1}$ were much larger than $\delta_{\gamma 1}$, then the Higgs decay to $\gamma Z$ would dominate over its decay to $\gamma\gamma$, which would likely have been confirmed by the Higgs searches made thus far in the LHC. It is, therefore, safer to assume that $\delta_{\gamma Z1}$ isn't much larger than $\delta_{\gamma 1}$ and that the scale of new physics is likely to be significantly larger than $10$ TeV. Nonetheless, the scale of new physics from $h\gamma Z$ could still be lower than that from $h\gamma\gamma$ and thus within the reach of the $100$ TeV collider. For instance, a moderate value of $\delta_{\gamma Z1} \sim \delta_{\gamma 1} \simeq 0.1$ would put the scale of new physics at $\sim 48$ TeV, which is significantly less than the scale of new physics that could be obtained from the coupling $h\gamma\gamma$. Therefore, the $h\gamma Z$ coupling remains more promising for searching for physics BSM compared with the $h\gamma\gamma$ coupling. 

\begin{figure}[!t]
\centerline{\begin{minipage}{0.8\textwidth}
\centerline{\includegraphics[width=300pt]{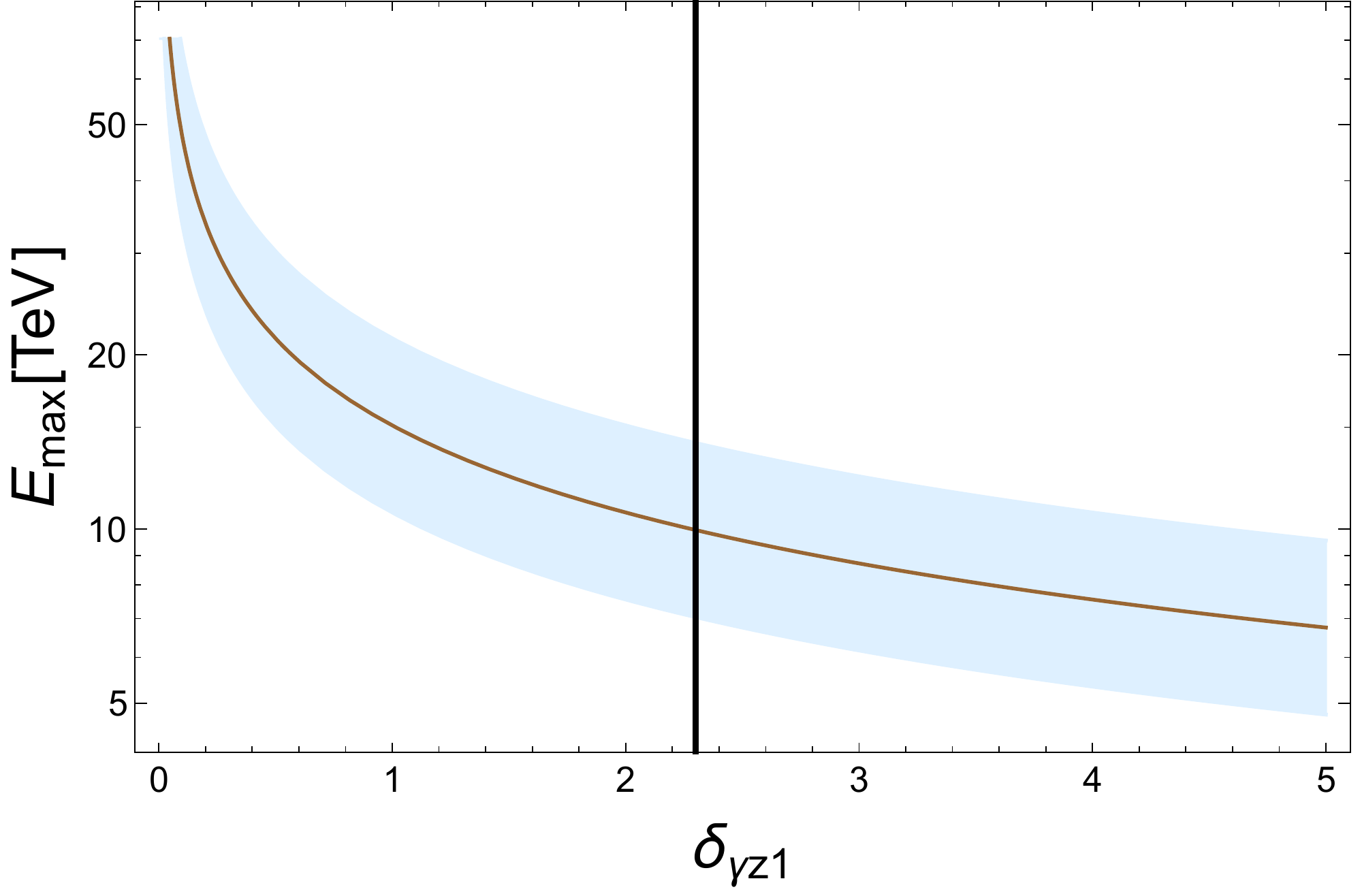}}
\caption{\small 
The unitarity bound as a function of $\delta_{\gamma Z1}$. The model-independent bound is obtained from eq. (\ref{eq:deltaZ1_strongest}). The band around the model-independent bound results from varying the unitarity bound to $\frac{1}{2} \leq \hat{\mathscr{M}} \leq 2$. The solid black line represents the $95 \%$ C.L. limits on $\delta_{\gamma Z1}$ from ATLAS \cite{ATLAS:2015egz}. The HL-LHC projections are unavailable.}
\label{fig8}
\end{minipage}}
\end{figure}

Figure \ref{fig8} shows the model-independent bound from eq. (\ref{eq:deltaZ1_strongest}) as a function of $\delta_{\gamma Z1}$, together with the $95 \%$ C.L ATLAS bound \cite{ATLAS:2015egz}. The HL-LHC projections for this coupling are unavailable.

\subsection{Model-Independent Bounds on the Scale of New Physics from the Couplings $h^{n}\gamma\gamma$}\label{Sec:hgZBound}
Similar to Section \ref{Sec:hggBoundn}, we can extend eq. (\ref{eq:HgaZ_Lag1}) to $n$-Higgses by introducing the operators
\begin{equation}\label{eq:nHgaZ_Lag1}
\mathcal{L}_{h^{n}\gamma Z} = A_{\mu\nu}Z_{T}^{\mu\nu} \sum_{n=1}^{\infty}\frac{c_{\gamma Zn}}{n!}\frac{\alpha}{\pi}\Big( \frac{h}{v} \Big)^{n},
\end{equation}
where the Wilson coefficients $c_{\gamma Zn}$ parametrize the couplings of $n$-Higgses to a photon and a $Z$ boson after integrating out the top and the $W$ loops as usual. Similar to the case of $\gamma\gamma$, only $c^{\text{SM}}_{\gamma Z1}$ has been calculated, thus it is more convenient to use it to define the deviations in $c_{\gamma Zn}$
\begin{equation}\label{eq:nHgaZ_dev}
\delta_{\gamma Zn} \equiv \frac{c_{\gamma Zn}-c^{\text{SM}}_{\gamma Zn}}{c^{\text{SM}}_{\gamma Z1}},
\end{equation}
with $\delta_{\gamma Zn} \rightarrow 0$ corresponding to the SM limit. Restoring the Goldstones, and only keeping the transverse mode of $Z$, we can expand eq. (\ref{eq:nHgaZ_Lag1}) as\footnote{Notice that eq. (\ref{eq:nHgaZ_Lag2}) can be obtain from eq. (\ref{eq:nHgaga_Lag2}) by replacing one of the photons with the transverse mode of $Z$, $A^{\mu\nu} \rightarrow Z^{\mu\nu}_{T}$ and relabeling the deviations and Wilson coefficient appropriately.} 
\begin{dmath}\label{eq:nHgaZ_Lag2}
\mathcal{L}_{h^{n}\gamma Z} = \frac{\alpha c^{\text{SM} }_{\gamma Z1}}{2\pi} A_{\mu\nu}Z_{T}^{\mu\nu} \Big\{ \Big(\frac{\delta_{\gamma Z1}}{v^{2}} \Big) \Big[Z_{L}^{2} + 2 W_{L}^{+}W_{L}^{-} \Big] + \Big(\frac{\delta_{\gamma Z2}}{v^{2}} \Big)h^{2} + 
 \Big( \frac{\delta_{\gamma Z2}-\delta_{\gamma Z1}}{v^{3}}\Big)h \Big[Z_{L}^{2} + 2 W_{L}^{+}W_{L}^{-} \Big]  + \Big(\frac{\delta_{\gamma Z3}}{3v^{3}} \Big)h^{3} + 
 \Big( \frac{\delta_{\gamma Z2}- \delta_{\gamma Z1}}{4v^{4}}\Big) \Big[Z_{L}^{2} + 2 W_{L}^{+}W_{L}^{-} \Big]^{2}+ 
 \Big( \frac{\delta_{\gamma Z1}-\delta_{\gamma Z2}+\frac{1}{2}\delta_{\gamma Z3}}{v^{4}}\Big) h^{2} \Big[  Z_{L}^{2} + 2 W_{L}^{+}W_{L}^{-}\Big] + \Big(\frac{\delta_{\gamma Z4}}{12v^{4}} \Big)h^{4} + \dots 
\Big\},
\end{dmath}
and as usual, eq. (\ref{eq:nHgaZ_Lag2}) also includes the operators in eq. (\ref{eq:nHgaZ_Lag1}). The strongest model-independent bounds up to the dim-8 operator shown above are given by

\begin{dmath}\label{eq:nHgaZ_Strongest}
\begin{aligned}
\gamma_{\pm}  Z_{T\pm}\rightarrow hh & :  \hspace{2mm} E_{\text{max}} \simeq \frac{18 \hspace{1mm}\text{TeV}}{|\delta_{\gamma Z2}|^{1/2}}, \\
\gamma_{\pm}Z_{T\pm} \rightarrow hhh & : \hspace{2mm}E_{\text{max}} \simeq \frac{13.5 \hspace{1mm}\text{TeV}}{|\delta_{\gamma Z3}|^{1/3}}, \\
\gamma_{\pm} Z_{T\pm} h \rightarrow hhh & : \hspace{2mm} E_{\text{max}} \simeq \frac{13.4 \hspace{1mm}\text{TeV}}{|\delta_{\gamma Z4}|^{1/4}}. \\
\end{aligned}
\end{dmath}

Similar to the case of $h^{n}\gamma\gamma$, it's easy to see that all processes involving any number of pure Higgses coupled to $\gamma$ and $Z_{T}$ are model-independent. Considering the operator
\begin{equation}\label{eq:2n+2_model_indep_Z}
\delta \mathcal{L}^{\gamma Z}_{\text{2n+2}} = \Big( \frac{c^{\text{SM}}_{\gamma Z1}\delta_{\gamma Z2n}\alpha}{(2n)!\pi}\Big) \Big( \frac{h}{v}\Big)^{2n} A_{\mu\nu}Z_{T}^{\mu\nu}.
\end{equation}
and extracting the amplitudes of all symmetric processes, we find
\begin{dmath}\label{eq:2n_bounds_Z}
\begin{aligned}
\gamma_{\pm} Z_{T\pm} h^{n-1}\rightarrow h^{n+1} & :  \hspace{2mm} E_{\text{max}} \simeq 4\pi v \Bigg( \frac{(n+1)!n!\sqrt{(n+1)!(n-1)!}}{4 \alpha c^{\text{SM}}_{\gamma Z1}|\delta_{\gamma Z2n}|}\Bigg)^{\frac{1}{2n}}, \\
\gamma_{\pm}h^{n}\rightarrow Z_{T\mp}  h^{n} & :  \hspace{2mm} E_{\text{max}} \simeq 4\pi v \Bigg( \frac{(n+1)!(n+1)!(n-1)!}{4\alpha c^{\text{SM}}_{\gamma Z1} |\delta_{\gamma Z2n}|}\Bigg)^{\frac{1}{2n}}. \\
\end{aligned}
\end{dmath}

\begin{figure}[!t]
\centerline{\begin{minipage}{0.8\textwidth}
\centerline{\includegraphics[width=250pt]{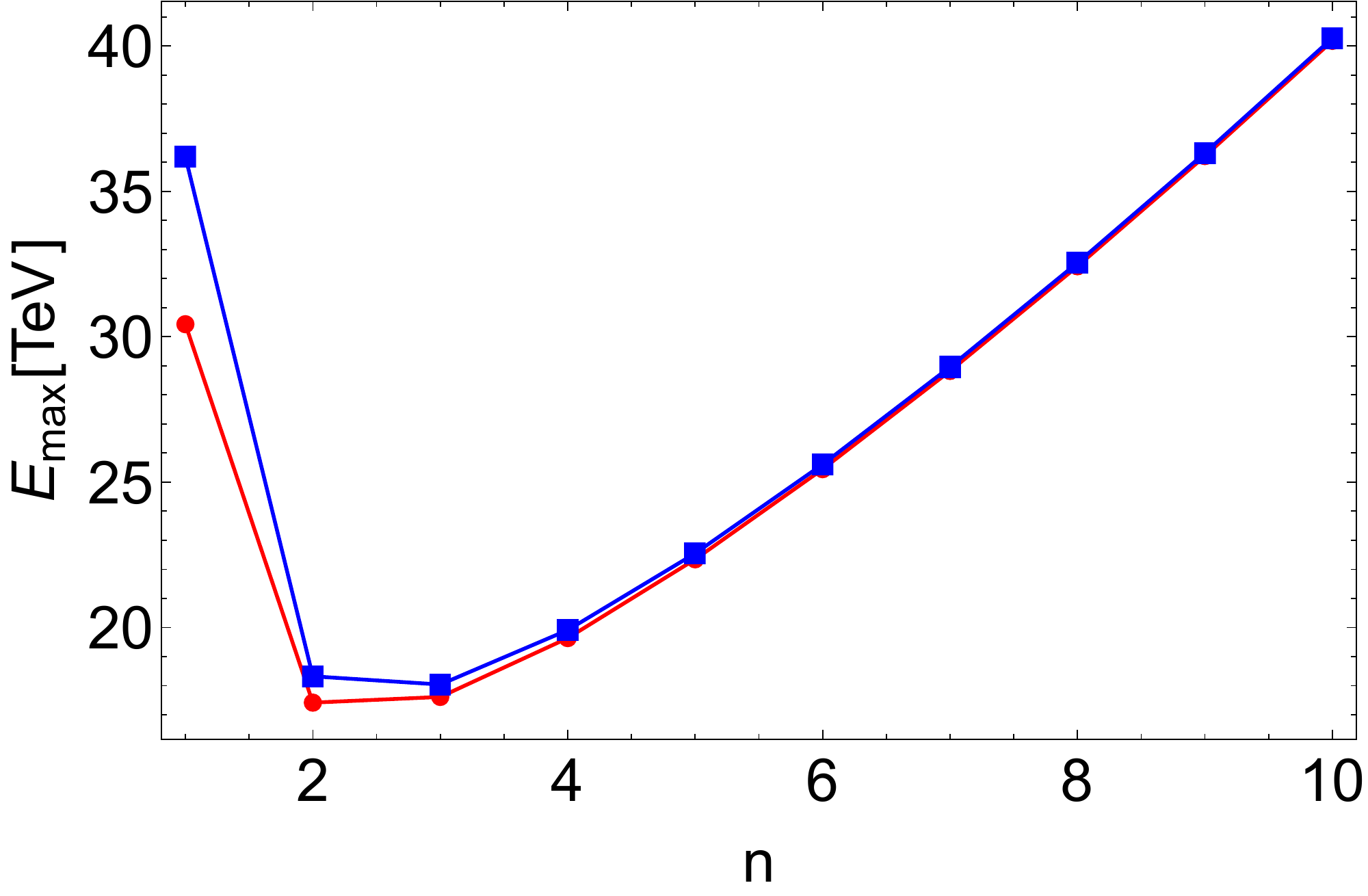}}
\caption{\small 
The unitarity-violating scale as a function on $n$ for $\gamma_{\pm} h^{n} \rightarrow Z_{T\mp} h^{n}$ (blue), and $\gamma_{\pm} Z_{T\pm} h^{n-1} \rightarrow h^{n+1}$ (red). The latter processes are dominant for all $n$. Here $\delta_{\gamma Z2n}$ is set to $1$.}
\label{fig9}
\end{minipage}}
\end{figure}

In Figure \ref{fig9}, we show a comparison between the two processes given in eq. (\ref{eq:2n_bounds_Z}), where we can see that unlike the $\gamma\gamma$ sector, here $\gamma_{\pm} Z_{T\pm} h^{n-1} \rightarrow h^{n+1}$ dominates for all $n$. This stems purely from combinatorics due to gauge bosons no longer being identical. We also observe that the same behavior found in the $\gamma\gamma$ sector extends here, namely that the strongest bound occurs at small $n$, and that it is not sensitive to $\delta_{\gamma Z2n}$, as long as they get smaller with $n$. This leads us to believe that the bounds in eq. (\ref{eq:deltaZ1_strongest}) and eq. (\ref{eq:2n_bounds_Z}) are not far from optimal as well.

As in the case with the $\gamma\gamma$ sector, the scale of new physics is beyond the LHC, unless the deviations are unnaturally large. However, it is still well within the reach of the $100$-TeV collider. For instance, for $\delta_{\gamma 3,4} = 0.1$, the scale of new physics can be $\lesssim 30$ TeV. In addition, the LHC and HL-LHC can still constrain $\delta_{\gamma Z2n}$ indirectly through precision measurements, as we show below.
% ---------------------------------------------------------------------------------
\subsection{Other Bounds}\label{Sec:hgZMixedProcesses}
Here we investigate the scale of new physics from processes that are not model-independent, limiting ourselves to the ones that only depend on $\delta_{\gamma Z1}$ and $\delta_{\gamma Z2}$. The strongest bounds originate from
\begin{dmath}\label{eq:NLObounds_Z}
\begin{aligned}
\gamma_{\pm}Z_{T\pm} \rightarrow W_{L}^{+}W_{L}^{-} h & :  \hspace{2mm} E_{\text{max}} \simeq \frac{10 \hspace{1mm}\text{TeV}}{|\delta_{\gamma Z2} - \delta_{\gamma Z1}|^{1/3}}, \\
\gamma_{\pm} Z_{T\pm} W_{L}^{+} \rightarrow  W_{L}^{+}W_{L}^{-}W_{L}^{+}  & : \hspace{2mm} E_{\text{max}} \simeq \frac{9.8 \hspace{1mm}\text{TeV}}{|\delta_{\gamma Z2} - \delta_{\gamma Z1}|^{1/4}}. \\
\end{aligned}
\end{dmath}

\begin{figure}[!t]
\centerline{\begin{minipage}{0.8\textwidth}
\centerline{\includegraphics[width=300pt]{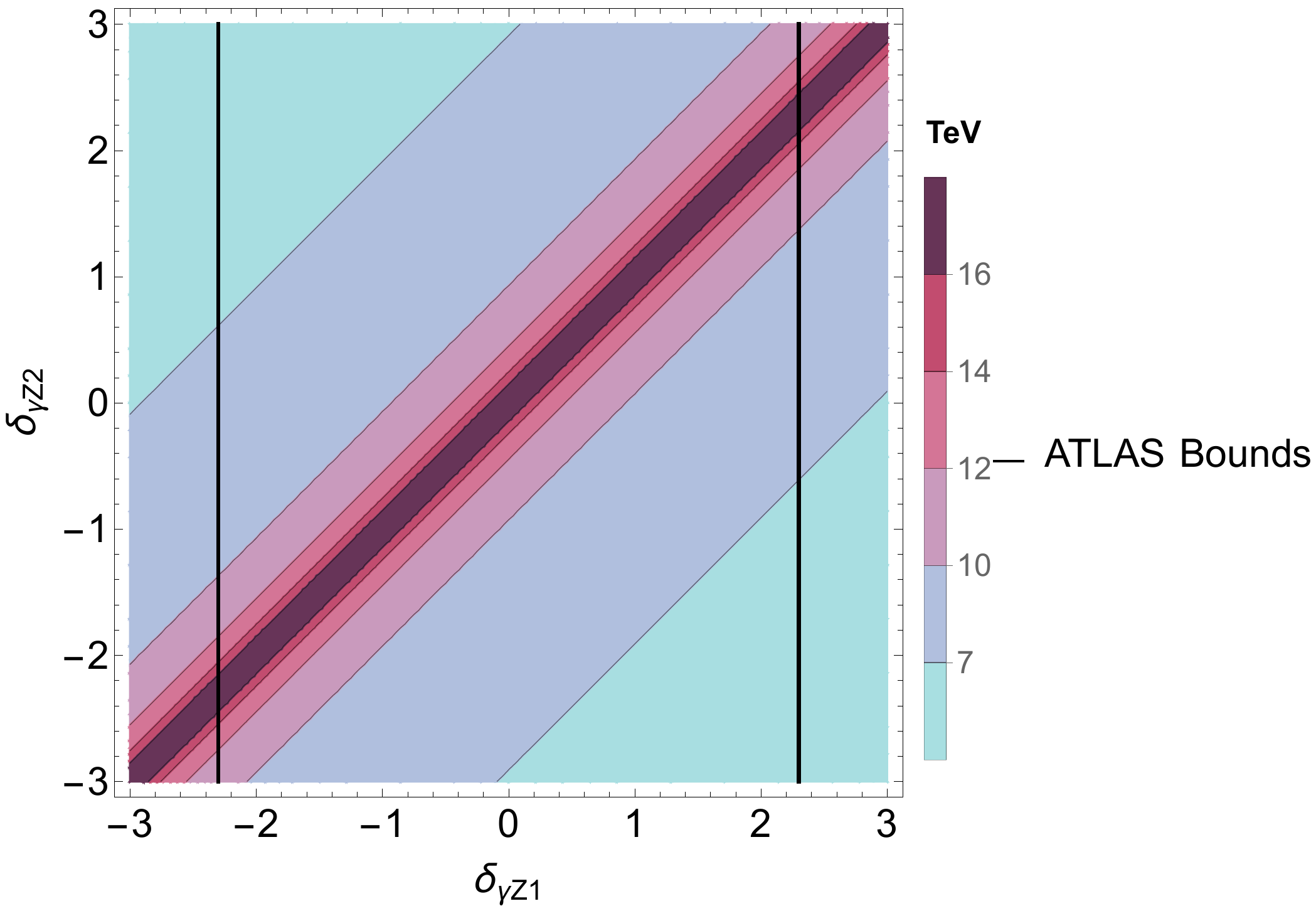}}
\caption{\small 
The unitarity-violating scale that depends on both $\delta_{\gamma Z1}$ and $\delta_{\gamma Z2}$. The solid black line represents the $95 \%$ C.L. limits on $\delta_{\gamma Z1}$ from ATLAS \cite{ATLAS:2015egz}. HL-LHC projections are unavailable.}
\label{fig10}
\end{minipage}}
\end{figure}

We show the strongest unitarity-violating scale from eq. (\ref{eq:NLObounds_Z}) in Figure \ref{fig10}, together with the $95 \%$ C.L. limits from ATLAS \cite{ATLAS:2015egz}. The diagonal corresponds to the vanishing of the corresponding operators in eq. (\ref{eq:nHgaZ_Lag2}), which indicates that higher-order operators should be included. The plot shows that the scale of new physics could be as low as $\sim 7$ TeV, which is within the reach of the LHC, however, this could only be the case if the deviations are somewhat large.
% ---------------------------------------------------------------------------------
\subsection{SMEFT Predictions from Unitarity}
\label{Sec:hgZSMEFT}

Similar to the $\gamma\gamma$ sector, we can use unitarity arguments to estimate the accuracy of SMEFT. For the $\gamma Z$ sector, we consider the SM plus the following dim-6 SMEFT operator

\begin{equation}\label{eq:gaZ_dim6_SMEFT}
\mathcal{L}^{\text{dim-6}}_{\text{SMEFT}} = \frac{1}{M^{2}}\Big( H^{\dagger}H - \frac{v^{2}}{2}\Big) A_{\mu\nu}Z^{\mu\nu},
\end{equation}
and demand that it dominates over higher-order operators. This yields the following predictions
\begin{equation}\label{eq:gaZ_SMEFTpred}
\delta_{\gamma Z1} = \frac{\pi v^{2}}{\alpha c^{\text{SM}}_{\gamma Z1}M^{2}}, \hspace{10mm} \delta_{\gamma Z2} = \delta_{\gamma Z1}, \hspace{10mm} \delta_{\gamma Zn} = 0 \hspace{2 mm} \text{for} \hspace{2mm} n  \geq 3.
\end{equation}

The requirement that the scale of new physics that comes from $\delta_{\gamma Z2}$ does not fall below the one that comes from  $\delta_{\gamma Z1}$, requires that the deviation of $\delta_{\gamma Z2}$ from the SMEFT prediction be small. Therefore we define
\begin{equation}\label{eq:gaZ_epsilon}
\epsilon_{\gamma 2Z} \equiv \frac{\delta_{\gamma Z2} - \delta^{\text{SMEFT}}_{\gamma Z2}}{\delta^{\text{SMEFT}}_{\gamma Z2}} \ll 1.
\end{equation}

\begin{figure}[!t]
\centerline{\begin{minipage}{0.8\textwidth}
\centerline{\includegraphics[width=300pt]{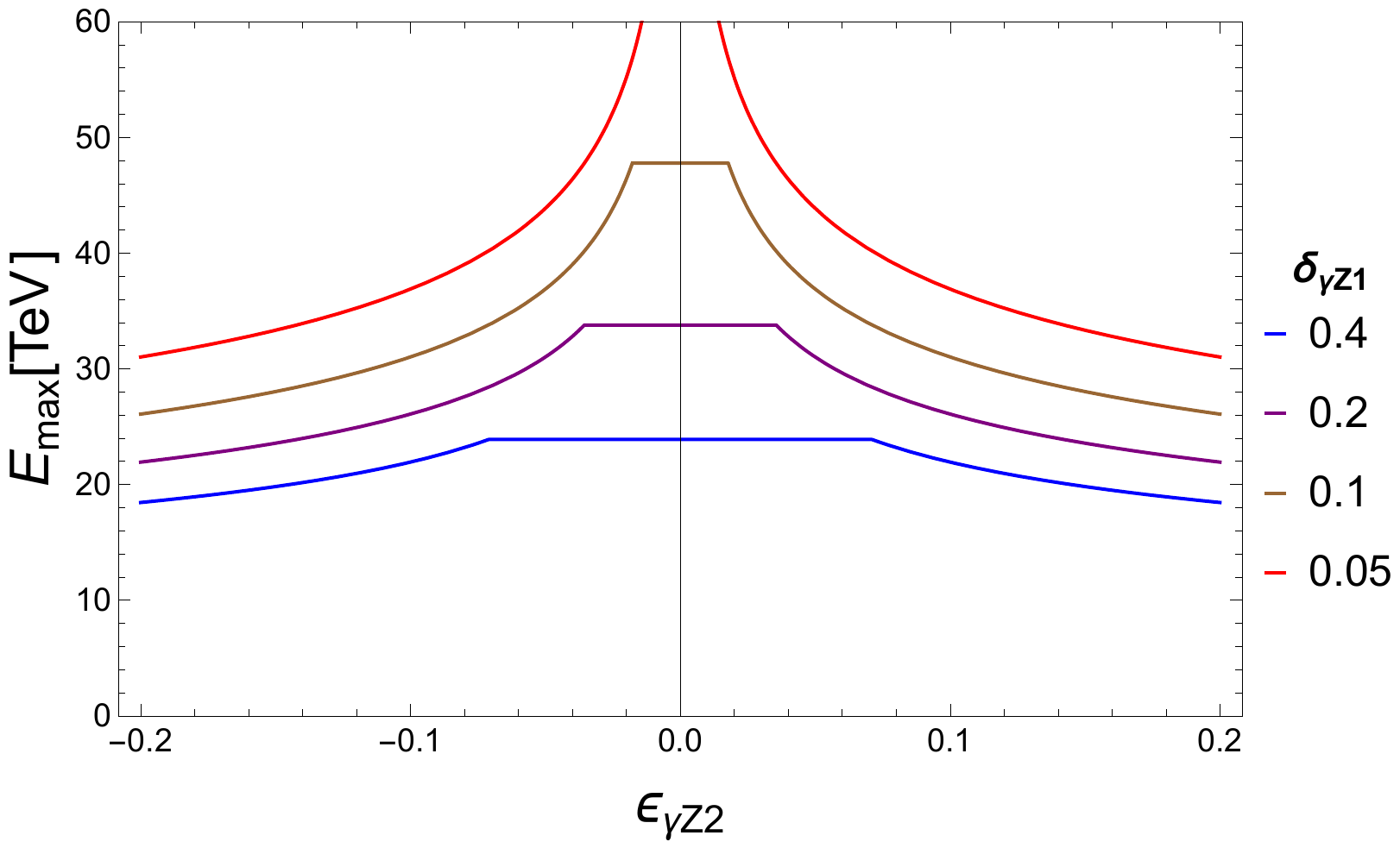}}
\caption{\small 
The unitarity-violating scale from processes that depend on both $\delta_{\gamma Z1}$ and $\delta_{\gamma Z2}$ as function of the fractional deviation $\epsilon_{\gamma Z2}$ from the predictions of the dim-6 operator. See eqs. (\ref{eq:gaZ_SMEFTpred}) and (\ref{eq:gaZ_epsilon}).}
\label{fig11}
\end{minipage}}
\end{figure}

The processes that are most sensitive to new physics are the ones given in eq. (\ref{eq:NLObounds_Z}). We plot the bound from this process against the fractional deviation from the SMEFT predictions $\epsilon_{\gamma Z2}$ in Figure \ref{fig11}, where the plot shows the usual behavior of improved accuracy of SMEFT as the scale of new physics increases. For example, if $|\delta_{\gamma Z1}|$ is constrained to be $\lesssim 0.1$, then this would push the scale of new physics to be $\gtrsim 48$ TeV, which would, in turn, constrain $|\delta_{\gamma Z1}|$ to be within $2\%$ of the SMEFT prediction in eq. (\ref{eq:gaZ_SMEFTpred}).

\section{Electroweak Precision Observables}\label{sec:EWPO}
In addition to leading to unitarity violation at tree-level, any deviation in the Higgs couplings compared with the SM predictions will contribute to the EWPO. Therefore, it is worthwhile investigating whether EWPO could yield bounds on these deviations that are stronger than the collider bounds found in the literature. In this section, we study the bound on $\delta_{\gamma n}$ and $\delta_{\gamma Z n}$ that arise from the oblique parameters. Recall that the oblique parameters are given by
\begin{equation}\label{eq:STU}
\begin{aligned}
T & \equiv  \frac{1}{\alpha} \Big[ \frac{\Pi_{WW}(0)}{m_{W}^{2}} - \frac{\Pi_{ZZ}(0)}{m_{Z}^{2}} \Big],\\
S & \equiv  \frac{4c^{2}s^{2}}{\alpha} \Big[ \frac{\Pi_{ZZ}(m_{Z}^{2})-\Pi_{ZZ}(0)}{m_{Z}^{2}} - \Big( \frac{c^{2}-s^{2}}{c s}\Big) \frac{\Pi_{\gamma Z}(m_{Z}^{2})}{m_{Z}^{2}} - \frac{\Pi_{\gamma \gamma}(m_{Z}^{2})}{m_{Z}^{2}} \Big],\\
U  & \equiv \frac{4 s^{2}}{\alpha} \Big[ \frac{\Pi_{WW}(m_{W}^{2})-\Pi_{WW}(0)}{m_{W}^{2}} - \frac{c}{s} \frac{\Pi_{\gamma Z}(m_{Z}^{2})}{m_{Z}^{2}} - \frac{\Pi_{\gamma \gamma}(m_{Z}^{2})}{m_{Z}^{2}} \Big] - S.
\end{aligned}
\end{equation}
where $\Pi_{XX}$ are the BSM contributions to the $XX$ propagator, and $s$ and $c$ are the sine and the cosine of the Weinberg angle respectively. Inspecting eq. (\ref{eq:effLag1}), (\ref{eq:nHgaga_Lag1}) and (\ref{eq:nHgaZ_Lag1}), it is easy to see that no model-independent bounds can be obtained from the $STU$ parameters, as each parameter receives several contributions that arise from multiple deviations. Hence it is not possible to set limits on one particular deviation. However, it could be possible to set model-dependent limits on the deviations if we consider only one deviation at a time and set all others to zero. 

As an illustrative example, we consider the corrections associated with $\delta_{\gamma 2}$ and $\delta_{\gamma Z 2}$. These deviations give rise to corrections to $\Pi_{\gamma\gamma}$ and $\Pi_{\gamma Z}$, which arise from Higgs loops with $hh\gamma\gamma$ and $hh\gamma Z$ vertices. These loops can easily be evaluated. In the $\overline{MS}$ scheme, they are given by
\begin{eqnarray}\label{eq:STU_loops1}
\Pi_{\gamma\gamma}(p^{2}) = \delta_{\gamma 2} \Big( \frac{\alpha m_{h}^{2}}{16\pi^{3} v^{2}c^{\text{SM}}_{\gamma 1}}\Big) p^{2} \log{\Big( \frac{m_{h}^{2}}{\mu^{2}} \Big)},\\
\Pi_{\gamma Z}(p^{2}) = \delta_{\gamma Z2} \Big( \frac{\alpha m_{h}^{2}}{16\pi^{3} v^{2}c^{\text{SM}}_{\gamma Z 1}}\Big) p^{2} \log{\Big( \frac{m_{h}^{2}}{\mu^{2}} \Big)}.\label{eq:STU_loops2}
\end{eqnarray}

Before discussing the effects on the oblique parameters, we should note that in our approach, the lack of an expansion scale leads to ambiguity at the quantum level. This ambiguity is manifest in the dependence of eqs. (\ref{eq:STU_loops1}) and (\ref{eq:STU_loops2}) on the renormalization scale $\mu$, and does not arise in EFTs with a well-defined expansion scale such as in SMEFT. The explanation is rather simple; when parametrizing the BSM physics as the deviations shown in eq. (\ref{eq:effLag1}), the theory is UV-incomplete and the loop amplitudes are UV-divergent. In order to extract calculable results, one has to assume that the UV divergence can be subtracted by an appropriate counter-term in dimensional regularization, that is assumed to arise from the renormalizable full theory. As the UV theory is unknown, a dependence on the renormalization scale $\mu$ remains in the final amplitude. Nonetheless, as this dependence on the renormalization scale $\mu$ is only logarithmic in dimensional regularization, our results are not invalidated.

It is easy to see that only the $S$ and $T$ parameters are affected. Given that both parameters are constrained at $O(0.1)$, it is easy to show that the bounds on $\delta_{\gamma 2}$ and $\delta_{\gamma Z2}$ obtained from these parameters are negligible. For example, the $S$ parameter constrains $\delta_{\gamma 2} > 375$, where we have set $\mu^{2} =m_{Z}^{2}$. This is much weaker than collider bounds. The same behavior is expected to carry out for all other deviations. Therefore, we can safely conclude that the bounds obtained from the EWPO do not compete with collider bounds.

\section{Possible UV Completions}\label{sec:UVcompletion}
In this section, we present a few simple extensions to the SM and show how our EFT approach is mapped to the UV completions. In particular, we consider the 2-Higgs Doublet Model (2HDM) - Type II, the scalar singlet extension of the Higgs sector, and the extension of the SM with a top quark partner. In each scenario, we briefly summarize the UV model, show how the leading deviations map into the UV parameters and calculate the unitarity-violating scale given the experimental constraints on the model's parameters.
\subsection{The 2HDM - Type II}
In the 2HDM, the Higgs sector is extended to 2 doublets: an up-type doublet and a down-type doublet, each developing its own VEV
\begin{equation}\label{eq:2HDM}
H_{u} = \begin{pmatrix}
H^{+}_{u}\\
\frac{v_{u}+H^{0}_{u}}{\sqrt{2}}
\end{pmatrix}, \hspace{1cm}
H_{d} = \begin{pmatrix}
\frac{v_{d}+H^{0}_{d}}{\sqrt{2}}\\
H^{-}_{d}\\
\end{pmatrix},
\end{equation}
where $\tan{\beta} \equiv v_{u}/v_{d}$, $\frac{\pi}{2} > \beta > 0$. The 2HDM has 8 degrees of freedom, 3 of which are eaten by the $W^{\pm}$ and $Z$, whereas the rest give rise to 2 $CP$-even neutral Higgses $h$ and $H$, 1 $CP$-odd neutral Higgs $A$, and two charged Higges $H^{\pm}$. The physical $CP$-even neutral Higgses are obtain via the rotation
\begin{equation}\label{eq:2HDMRot}
\begin{pmatrix} 
H\\
h
\end{pmatrix} = 
\begin{pmatrix} 
\cos{\alpha} && \sin{\alpha}\\
-\sin{\alpha} && \cos{\alpha}
\end{pmatrix} 
\begin{pmatrix} 
\text{Re}(H^{0}_{d})\\
\text{Re}(H^{0}_{u})
\end{pmatrix},
\end{equation}
where $h$ is the SM Higgs. In type II, $H_{u}$ couples to the up-type quarks, whereas $H_{d}$ couples to down-type quarks and leptons. This implies that the coupling of $h$ to the top quark gets multiplied by $\frac{\cos{\alpha}}{\sin{\beta}}$, whereas the coupling to the $W$ gets modified by $\sin{(\beta -\alpha )}$. Therefore, the values of $c^{\text{2HDM}}_{\gamma 1}$ and $c^{\text{2HDM}}_{\gamma Z1}$ which result from integrating out the top and $W$ loops are obtained by simply modifying the SM contributions accordingly, and $\delta_{\gamma 1}$ and $\delta_{\gamma Z1}$ are obtained from eqs. (\ref{eq:dev1}) and (\ref{eq:hgaZ_dev}) respectively. Given the bounds on $\alpha$ and $\beta$ in the 2HDM - Type II \cite{Atkinson:2021eox}, the unitarity-violating scale is found to be
\begin{equation}\label{eq:2HDM_Emax}
E_{\text{max}}^{\gamma\gamma (\gamma Z)} \gtrsim 143 \hspace{2mm}(198) \hspace{1mm} \text{TeV}.
\end{equation}

This is much larger than the best fit obtained from \cite{Atkinson:2021eox}, which predicts masses of heavy Higgses of around $2$ TeV. This is due to the smallness of $\delta_{\gamma1}$ and $\delta_{\gamma Z1}$ in this scenario, which makes the bounds weak. This is to be expected since these couplings are loop-induced.
\subsection{A Scalar Singlet Extension}
We consider extending the SM Higgs sector with a scalar singlet $\Phi$ that develops a VEV $v_{1}$. In the unitary gauge, the SM scalar sector becomes
\begin{equation}\label{eq:SMH}
H = \begin{pmatrix}
0\\
\frac{v_{0}+\phi}{\sqrt{2}}
\end{pmatrix}, \hspace{1cm}
\Phi = v_{1} + \chi.
\end{equation}
where $v_{0}$ is the SM Higgs VEV and $\phi$ and $\chi$ are the gauge eigenstates of the SM Higgs and scalar singlet respectively. The part of the Lagrangian that is relevant for our study is given by
\begin{equation}\label{eq:Singlet_Lag}
\mathcal{L} = (D_{\mu}H)^{\dagger}(D^{\mu}H) + \frac{1}{2}(\partial_{\mu} \Phi)(\partial^{\mu}\Phi) + V(H,\Phi) -\big( y_{t} \bar{t}_{L}\tilde{H}t_{R} + h.c.\big),
\end{equation}
where $\tilde{H} = \epsilon H^{*}$, $y_{t}$ is the SM top quark Yukawa coupling, and the scalar potential is given by
\begin{equation}\label{eq:ScalarV}
V(H,\Phi) = m_{H}^{2}(H^{\dagger}H) +\frac{1}{2} m_{\Phi}^{2}\Phi^{2} + \lambda (H^{\dagger}H)^{2} - \mu \Phi (H^{\dagger}H).
\end{equation}

The potential in eq. (\ref{eq:ScalarV}) is minimized by imposing $\partial V/\partial v_{0} = \partial V/\partial v_{1} = 0$, and the mass matrix is obtained by substituting eq. (\ref{eq:SMH}) in $V(H,\Phi)$. The masses of the physical SM Higgs $h$ and heavy Higgs $S$ are obtained through diagonalizing the mass matrix, which gives the following mass eigenvalues for the SM Higgs $h$ and the heavy Higgs $S$
\begin{equation}\label{eq:HiggsMasses}
M_{h,S}^{2} = \frac{1}{2}m_{\Phi}^{2} + \lambda v_{0}^{2} \mp \frac{1}{2} \sqrt{(m_{\Phi}^{2}-2\lambda v_{0}^{2})^{2}+4\mu^{2}v_{0}^{2}}.
\end{equation}

Notice that in the limit $\mu \rightarrow 0$, the heavy Higgs $S$ decouples and the SM limit is restored. The mass eigenstates $h$ and $S$ can be obtained from the gauge eigenstates $\phi$ and $\chi$ via the rotation
\begin{equation}\label{eq:SingletRot}
\begin{pmatrix} 
h\\
S
\end{pmatrix} = 
\begin{pmatrix} 
\cos{\alpha} && \sin{\alpha}\\
-\sin{\alpha} && \cos{\alpha}
\end{pmatrix} 
\begin{pmatrix} 
\phi\\
\chi
\end{pmatrix},
\end{equation}
where
\begin{eqnarray}\label{eq:alpha_Singlet}
\sin{2\alpha} = \frac{2 \mu v_{0}}{M_{S}^{2}-M_{h}^{2}},\\
\tan{\alpha} = \frac{\mu v_{0}}{M_{h}^{2}-2\lambda v_{0}^{2}}.
\end{eqnarray}
A straightforward calculation shows that the couplings of the SM Higgs to the top quark and $W$ boson get modified by $\cos{\alpha}$. Therefore, both $c_{\gamma 1}$ and $c_{\gamma Z1}$ in this scenario are simply multiplied by $\cos{\alpha}$, and the deviations $\delta_{\gamma (Z) 1} = \cos{\alpha} -1$. The constraints on $\alpha$ are obtained from the Higgs signal strength \cite{ATLAS:2019nkf}
\begin{equation}\label{eq:SingletBound}
\cos^{2}{(\alpha)} \in [0.95,1],
\end{equation}
and with this bound, the unitarity-violating scale in this scenario is found to be
\begin{equation}\label{eq:Singlet_Emax}
E_{\text{max}}^{\gamma\gamma (\gamma Z)} \gtrsim 150 \hspace{2mm}(95) \hspace{1mm} \text{TeV}.
\end{equation}

Once again, and for the same reasons as in the case of the 2HDM, the bounds are much larger than the best fit predicted for the mass of the heavy Higgs, which is predicted to be $\lesssim 1$ TeV \cite{Godunov:2015nea}. 
\subsection{Single Top Partner} 
Finally, we consider extending the SM with a single top quark partner $T$, that couples to the SM Higgs via the interaction
\begin{equation}\label{eq:topPartnet}
\mathcal{L}_{\text{int}} = - \frac{m_{T}}{v}h\overline{T}T.
\end{equation}

Assuming that the top partner is a color triplet with $Q = 2/3$, then the Higgs couplings to $\gamma\gamma$ and $\gamma Z$ will receive an additional contribution from the fermion loop in Figure \ref{fig2} with $T$ running in the loop. This additional contribution is simply found by substituting the mass of the top with $m_{T}$ in the fermion loop functions in the $\gamma\gamma$ and $\gamma Z$ sectors. Hence the deviations are readily found
\begin{equation}
\delta_{\gamma x1} = \frac{c_{\gamma x1}^{\text{SM}}(m_{t} \rightarrow m_{T}, g_{W} \rightarrow0)}{c_{\gamma x1}^{\text{SM}}}.
\end{equation}
where $x = \gamma, Z$. The fermion loop functions saturate quickly for large arguments, and since $m_{T}$ is constrained to be $\gg m_{t}$, the unitarity-violating scale quickly saturates, yielding the following bound which is independent of $m_{T}$
\begin{equation}\label{eq:topPartner_Emax}
E_{\text{max}}^{\gamma\gamma (\gamma Z)} \gtrsim 45 \hspace{2mm}(64) \hspace{1mm} \text{TeV}
\end{equation}

A top quark partner is predicted by many SM extensions, such as extra dimensions and composite Higgs models. Most of these models predict a top partner mass of a few hundred GeV to a few TeV (see for instance \cite{Cacciapaglia:2018qep}), which is much lower than the limits we have.

In all scenarios, we found the scale of new physics as predicted by unitarity arguments, is much larger than what is predicted for these models. This is to be expected as unitarity arguments usually give bounds for new physics that are much larger than the actual scale (for example, the unitarity bound on the mass of the SM Higgs was predicted to be $1$ TeV, compared to the actual mass which was discovered at $125$ GeV). In addition, the nature of the $h\gamma\gamma$ and $h\gamma Z$ couplings being loop-induced, makes them too small, which in turn leads to high bounds. One, in general, would expect lower bounds in these UV scenarios by studying tree-level couplings, like $hWW$ for instance.
% =========================================================================================
% =========================================================================================
\section{Conclusions}
\label{sec:conclusions}
In this paper, we investigated the scale of unitarity violation that arises from modifications to the Higgs couplings to $\gamma\gamma$ and $\gamma Z$. The unitarity of the SM at high energy relies on delicate cancelations among the various higher-order operators. This implies that any deviation in these couplings from the SM predictions would upset these cancelations, which in turn would lead to processes that have amplitudes that grow with energy, which would eventually violate unitarity at some high energy scale, signaling the onset of new physics.

In this paper, we focused on the couplings $h^{n}\gamma\gamma$ and $h^{n}\gamma Z$ with $n\geq 1$. In the SM, $h\gamma\gamma$ and $h\gamma Z$ are known theoretically to percent level, whereas experimentally they are only constrained to be $0.1$ and $2.3$ respectively. In addition, the other couplings with $n > 1$ were neither calculated theoretically nor measured experimentally. This leaves plenty of room for new physics BSM. Experiments in the future, including the HL-LHC and the future $100$ TeV collider, can help probe these couplings.

We found the unitarity-violating scale in the $h^{n}\gamma\gamma$ and $h^{n}\gamma Z$ to be larger than what was found in \cite{Chang:2019vez, Abu-Ajamieh:2020yqi} from the Higgs couplings to the top quark and $W/Z$, in addition to the Higgs trilinear coupling. This is to be expected given the fact that these couplings are loop-induced and thus are much weaker than the former tree-level ones. Specifically, we found that the current level of constraints on $h\gamma\gamma$ allows for new physics as high as $\sim 69$ TeV. On the other hand, since the couplings $h^{n}\gamma\gamma$ for $ n \geq 2$ are essentially unconstrained, the scale on new physics could be much lower. For instance, with the conservative assumption that $|\delta_{\gamma n}| \leq 1$, the scale of new physics could be as low as $\sim 12$ TeV. Similarly, the less stringent constraints on the coupling $h\gamma Z$ allow for the scale of new physics to be lower than the $\gamma\gamma$ sector and possibly below $10$ TeV. However, this is unlikely as it would require an unnaturally large $\delta_{\gamma Z1}$. Nonetheless, if we conservatively assume that $\delta_{\gamma Z1} \sim \delta_{\gamma1}$, then the scale of new physics would be $\sim 48$ TeV, which is still significantly lower than the $\gamma\gamma$ sector. Similarly, with the conservative assumption that $|\delta_{\gamma Zn}| \leq 1$, the scale of new physics could be as low as $\sim 7$ TeV, which might be within the reach of Run III of the LHC.

We also found that from the unitarity argument alone, we can both make quantitative statements about the accuracy of SMEFT, and indirectly set limits on the various couplings, especially the ones that are difficult to measure in colliders. For example, the HL-LHC is projected to measure the coupling $h\gamma\gamma$ at a $\pm 3.6\%$ level; if this coupling is found to conform to the SM predictions to that level, then the scale of new physics from this operator is pushed well above 100 TeV, which in turn will place stringent constraints on the coupling $h^{2}\gamma\gamma$, as it will be constrained to be within $\pm 0.2\%$ of the SMEFT predictions.

We discussed the EWPO and found that they do not provide stronger constraints on the couplings in consideration compared with the constraints from colliders. We also discussed a few simple UV completions and matched them to our EFT approach.

Our main conclusion is that the current level of measurement of the Higgs properties leaves ample room for new physics BSM, whether through direct or indirect searches. A completely model-independent bottom-up approach can help us probe the scale of new physics from unitarity considerations alone. We found that while the scale of new physics in the $h\gamma\gamma$ and $h\gamma Z$ is mostly beyond the reach of the LHC and HL-LHC, it is well within the reach of the future $100$ TeV collider. In addition, accurate measurements of the Higgs properties at lower energies can help both determine the scale of new physics and place stringent limits on its couplings. 
% =========================================================================================

\section*{Acknowledgments}
I would like to thank Spencer Chang for answering my questions. I also thank Markus Luty and Miranda Chan with whom the first part of this project was published. This work was supported by the C.V. Raman fellowship from CHEP at IISc.
%======================================================================
\appendix
\setcounter{section}{0}
\section{Calculation Techniques and Results}
\label{sec:appendix}
In this appendix, we show how we define the states used to calculate the unitarity bounds, explain how the bounds are calculated, and present our results. Some of the materials are repeated from \cite{Abu-Ajamieh:2020yqi} and are only presented here for the reader's convenience. In addition, here we also show how to include states with photons and compare our unitarity bounds with those obtained from partial wave unitarity.
\subsection{Scalar Amplitudes}
\label{app:A1}
We first discuss amplitudes involving only scalar fields, which include amplitudes with longitudinal $W$ and $Z$ bosons when we use the equivalence theorem.
Given $r$ species of scalars $\phi_{1}, \dots, \phi_{r}$; we define the states
\begin{dmath}\label{eq:ScalarState}
\begin{aligned}
\ket{P; k_{1},\dots, k_{r}} & \equiv C_{k_{1},\dots, k_{r}} \int d^{4}x e^{-i P.x} \phi_{1}^{(-)}(x)^{k_{1}} \dots \phi_{r}^{(-)}(x)^{k_{r}} \ket{0} \\ 
& = C_{k_{1},\dots, k_{r}}\int d\Phi_{k}(P;p_{1},\dots,p_{k}) \ket{\phi_{1}(p_{1}) \dots \phi_{r}(p_{k})}.
\end{aligned}
\end{dmath}

Here $k_{1}, \dots, k_{r}$ are non-negative integers that give the number of each species of particle in the state, $\phi_{i}^{(-)}$ is the negative frequency (creation operator) part of the interaction picture field $\phi_{i}$, $\ket{\phi_{1}(p_1) \cdots \phi_{r}(p_k)}$ 
is an ordinary $k$-particle state with $k = k_{1} + \cdots + k_{r}$, and
\begin{equation}\label{eq:PhaseSpace1}
d\Phi_{k}(P;p_{1},\dots,p_{k}) =\frac{d^{3}\vec{p}_{1}}{(2\pi)^{3}}\frac{1}{2E_{1}} \dots \frac{d^{3}\vec{p}_{k}}{(2\pi)^{3}}\frac{1}{2E_{k}} (2\pi)^{4}\delta^{4}(p_{1}+ \dots +p_{k} -P),
\end{equation}
is the Lorentz-invariant $k$-body phase space. These states are $s$-wave states defined by integrating $k$-particle states over the full phase space. The normalization of these states is chosen to be
\begin{equation}\label{eq:StateNormalization}
\braket{P';k'|P;k} = (2\pi)^{4}\delta^{4}(P'-P)\delta_{k'k},
\end{equation}
where we use the abbreviations
\begin{equation}\label{eq:abbr}
\ket{P;k} = \ket{P;k_{1}, \dots, k_{r}}, \hspace{5mm} \delta_{k'k} = \delta_{k'_{1}k_{1}} \dots \delta_{k'_{r}k_{r}},  \hspace{5mm} C_{k} = C_{	k_{1} \dots k_{r}}.
\end{equation}

The normalization constant is given by
\begin{equation}\label{eq:NormalizaionConst}
\frac{1}{|C_{k}|^{2}} = k_{1}! \dots k_{r}! \Phi_{k}(P),
\end{equation}
where
\begin{equation}\label{eq:PhaseSpace1}
\Phi_{k}(P) = \int d\Phi_{k}(P) = \frac{1}{8\pi(k-1)!(k-2)!}\Big(\frac{E}{4\pi}\Big)^{2k-4},
\end{equation}
is the total volume of phase space for massless particles with a center of mass energy $E = \sqrt{P^{2}}$. We then consider $S$-matrix elements between these states:
\begin{equation}\label{eq:TransferMat}
\bra{P';k'}T\ket{P;k} =(2\pi)^{4}\delta^{4}(P'-P)\hat{\mathscr{M}}(P;k_{1},\dots, k_{r} \rightarrow k'_{1},\dots, k'_{r} ),
\end{equation}
where $S = 1 + iT$. The amplitude $\hat{\mathscr{M}}$ is Lorentz-invariant and depends only on $P_{\mu}$, so it is a function of $E$ only. With the normalization in eq. (\ref{eq:StateNormalization}), unitarity of the $S$ matrix implies that these amplitudes satisfy
\begin{equation}\label{eq:UnitarityBound}
|\hat{\mathscr{M}}|	\hspace{1mm} \leq 1.
\end{equation}

For non-forward amplitudes, this follows directly from the unitarity of the $S$-matrix. For forward amplitudes ($k'_{i} = k_{i}$) a few additional steps are required to show that this holds for tree-level amplitudes, see \cite{Chang:2019vez}. This is the unitarity constraint we employ in this paper.

The Feynman rules for these amplitudes follow straightforwardly
from the standard rules. The result is that the amplitude $\hat{\mathscr{M}}$ are obtained from the standard Lorentz-invariant amplitude $\mathscr{M}$ by averaging over the initial and final state phase space:
\begin{equation}\label{eq:WeightedMat1}
\hat{\mathscr{M}}_{fi}(P) = C^{*}_{f}C_{i} \int d\Phi_{f}(P)d\Phi_{i}(P) \mathscr{M}_{fi},
\end{equation}
where $\mathscr{M}_{fi}$ is the usual Lorentz-invariant amplitude.
%\footnote{In more detail, eq. (\ref{eq:WeightedMat1}) is
%\begin{multline}\label{eq:WeightedMat2}
%\hat{\mathscr{M}}_{fi}(P;k_{1} \dots,k_{r} \rightarrow P';k'_{1} \dots,k'_{r} ) = C^{*}_{k'}C_{k} \int d\Phi_{k'}(P';p'_{1}, \dots, p'_{k'})d\Phi_{k}(P;p_{1}, \dots, p_{k}) \\
% \times \mathscr{M}(\phi_{1}(p_{1}) \dots \phi_{r}(p_{k})\rightarrow \phi_{1}(p'_{1}) \dots \phi_{r}(p'_{k'})).
%\end{multline}}

Because we are averaging over final state momenta, these amplitudes have
contributions from disconnected diagrams, with each disconnected component contributing an $\hat{\mathscr{M}}$ factor, leading to a form $\hat{\mathscr{M}} \propto \Pi_{i} \hat{\mathscr{M}}_i$ . However, the leading contribution to high-energy amplitudes always comes from connected diagrams.

In simple cases, these amplitudes can be computed in terms of the
total volume of phase space given in eq. (\ref{eq:PhaseSpace1}).
For example, for a single insertion of a coupling with no derivatives
we have
\begin{eqnarray}\label{eq:SingleInsertion}
\frac{\bra{P';k'} \int d^{4} x \phi_{1}^{n_{1}}(x) \dots \phi_{r}^{n_{r}}(x) \ket{P;k}}{(2\pi)^{4}\delta^{4}(P'-P)}  & = & C^{*}_{k'}C_{k} n_{1}! \dots  n_{r}! \Phi_{k'}(P')\Phi_{k}(P) \\
& = & \frac{1}{C_{k'}C^{*}_{k}} \frac{n_{1}! \dots n_{r}!}{k_{1}! \dots k_{r}!k'_{1}! \dots k'_{r}!},
\end{eqnarray}
where we assume $n_{i} = k_{i} + k'_{i}$. 
For diagrams with a single insertion of a vertex containing derivatives,
we use the identities
\begin{eqnarray}\label{eq:identies}
\int d\Phi_{k}(P; p_{1}, \dots, p_{k}) p_{1}^{\mu} & = & \frac{P^{\mu}}{k} \Phi_{k}(P), \\
\int d\Phi_{k}(P; p_{1}, \dots, p_{k}) p_{1}.p_{2} & = & \frac{P^{2}}{2{k \choose 2}} \Phi_{k}(P),
\end{eqnarray}
which hold for the case where all particles are massless.
\subsection{States with Photons and/or Transverse Modes}
\label{app:A2}
In order to consider states with photons and/or $Z_{T}$ modes, we write these fields as
\begin{eqnarray}\label{eq:1PhotonState1}
A_{\mu} (x) & = & \sum_{\lambda = 0}^{3} \int \frac{d^{3}\vec{p}}{(2\pi)^{3}2E_{p}} \Big[\epsilon_{\mu}(p,\lambda)e^{-i \vec{p}.\vec{x}} a_{p,\lambda} + \epsilon^{*}_{\mu}(p,\lambda)e^{i \vec{p}.\vec{x}} a^{\dagger}_{p,\lambda}\Big] \\
& \equiv & A_{\mu}^{(+)}(x) + A_{\mu}^{(-)}(x).
\end{eqnarray}
where $a^{\dagger}_{p,\lambda}$ and $a_{p,\lambda}$ are the creation and annihilation operators, and $\epsilon_{\mu}(p,\lambda)$ are the polarization vectors. For states with $k$ scalars of $r$ species and one photon or transverse mode, we write the state as
\begin{dmath}\label{eq:1PhotonState2}
\begin{aligned}
\ket{P; k_{1}, \dots, k_{r},q; \lambda} & \equiv C_{k_{1}, \dots, k_{r};1} \int d^{4}x e^{-iP .x} A_{\mu}^{(-)}(x) \phi_{1}^{(-)}(x)^{k_{1}} \dots \phi_{r}^{(-)}(x)^{k_{r}} \ket{0} \\ 
& = C_{k_{1}, \dots, k_{r};1} \int d\Phi_{k+1}(P;p_{1},\dots, p_{k}, q) \epsilon^{*}_{\mu}(q,\lambda) \ket{\phi_{1}(p_{1})\dots \phi_{r}(p_{k})A(q)}.
\end{aligned}
\end{dmath}

Imposing the normalization condition (\ref{eq:StateNormalization}), we find
\begin{eqnarray}\label{eq:1PhotonState3}
\frac{\braket{P',k',q';\lambda' |P,k,q;\lambda}}{(2\pi)^{4}\delta^{4}(P'-P)}  =  k_{1}! \dots k_{r}! |C_{k,1}|^{2} \int d\Phi_{k+1}(P;p_{1}, \dots, p_{k},q) \epsilon_{\mu}(q,\lambda) \epsilon^{*\mu}(q,\lambda ')
\end{eqnarray}
and we used the abbreviation in eq. (\ref{eq:abbr}). Since the polarization vector is normalized as
\begin{equation}\label{eq:PolNormalization}
\epsilon_{\mu}(q,\lambda) \epsilon^{*\mu}(q,\lambda ') = \delta_{\lambda \lambda '},
\end{equation}
it is easy to see that the normalization constant of a state with $k$ scalars and one photon (or transverse mode) is identical to that of a state with $k+1$ scalars
\begin{equation}\label{eq:1gaNormalization}
C_{k,1} = C_{k+1}.
\end{equation}

Similarly, states with k scalars and 2 photons can be expressed as
\begin{gather}\label{eq:1PhotonState2}
\begin{aligned}
\ket{P; k_{1}, \dots, k_{r}, q_{1}, q_{2}; \lambda_{1}\lambda_{2}}  &=
 C_{k_{1}, \dots, k_{r};2} \int d^{4}x e^{-iP .x} A_{1\mu}^{(-)}(x)  A_{2\nu}^{(-)}(x) \phi_{1}^{(-)}(x)^{k_{1}} \dots \phi_{r}^{(-)}(x)^{k_{r}} \ket{0} \\ 
&=C_{k_{1}, \dots, k_{r};2}  \int d\Phi_{k+1}(P;p_{1},\dots, p_{k}, q_{1},q_{2}) \epsilon^{*}_{\mu} (q_{1},\lambda_{1})\epsilon^{*}_{\nu} (q_{2},\lambda_{2}) \\
 & \hspace{21mm} \times \ket{\phi_{1}(p_{1})\dots \phi_{r}(p_{k})A_{1}(q_{1})A_{2}(q_{2})}.
\end{aligned}
\raisetag{14pt}
\end{gather}

Imposing the normalization condition in eq. (\ref{eq:StateNormalization}), in addition to the normalization of the polarization vectors in eq. (\ref{eq:PolNormalization}), it is easy to see that states with two photons can be treated in the same way as states with one photon, namely
\begin{equation}\label{eq:2gaNormalization}
C_{k,2} = C_{k+2}.
\end{equation}

Thus, for states with any number of photons, the normalization constants can be found using eqs. (\ref{eq:NormalizaionConst}) and (\ref{eq:PhaseSpace1}).

\subsection{Polarization Vectors Convention}
\label{app:A3}
Here we highlight the polarization vectors convention we used in this paper. For a photon propagating in the $\hat{z}$ direction, we define the polarization vector as
\begin{equation}\label{eq:Pol_z}
\epsilon_{\pm}^{\mu}(\hat{z}) = \frac{1}{\sqrt{2}}(0, -i, \mp 1, 0),
\end{equation}
where $- (+)$ corresponds to LH (RH) helicity. Notice that the polarization vectors are properly normalized. To obtain the polarization vectors in a general direction, where amplitudes have an azimuthal symmetry, we apply the rotation operator in 2D
\begin{equation}\label{eq:RotationOp}
\hat{n} = (0,0, \sin{\theta}, \cos{\theta}),
\end{equation}
where $\theta$ is the angle between $\hat{n}$ and $\hat{z}$. So, we get
\begin{equation}\label{eq:Pol_n}
\epsilon_{\pm}^{\mu}(\hat{n}) = \frac{1}{\sqrt{2}}(0, -i, \mp \cos{\theta}, \pm \sin{\theta}).
\end{equation}
\subsection{Sample Calculation}
\label{app:A4}
Here we illustrate how to calculate the amplitudes by presenting an explicit example. Consider the process $\gamma\gamma \rightarrow W_{L}^{+}W_{L}^{-}$, This process comes from the operator
\begin{equation}\label{eq:SampleOp}
\delta \mathcal{L}_{1} = \Bigg( \frac{c^{\text{SM}}_{\gamma1}\delta_{\gamma 1}\alpha}{\pi v^{2}} \Bigg) W_{L}^{+} W_{L}^{-} A_{\mu\nu}A^{\mu\nu}.
\end{equation}

Thus, the amplitude is given by
\begin{eqnarray}\label{eq:SampleCalc1}
\begin{aligned}
\hat{\mathscr{M}} = 2\Bigg( \frac{c^{\text{SM}}_{\gamma1}\delta_{\gamma 1}\alpha}{\pi v^{2}} \Bigg) C_{i}C^{*}_{f}\int d\Phi_{4}  \Big[ \bra{P';W_{L}^{+}W_{L}^{-}}W_{L}^{+}W_{L}^{-}\ket{0}\bra{0}(\partial_{\mu}A_{\nu})(\partial^{\mu}A^{\nu})\ket{P; \gamma\gamma} \\ 
 - \bra{P';W_{L}^{+}W_{L}^{-}}W_{L}^{+}W_{L}^{-}\ket{0}\bra{0}(\partial_{\mu}A_{\nu})(\partial^{\nu}A^{\mu})\ket{P; \gamma\gamma} \Big] \nonumber
 \end{aligned}
 \\= -2 . 2!  C_{i}C^{*}_{f}  \Bigg( \frac{c^{\text{SM}}_{\gamma1}\delta_{\gamma 1}\alpha}{\pi v^{2}} \Bigg) \int d\Phi_{4}  \Big[(p_{1}.p_{2})(\epsilon_{p_{1}}.\epsilon_{p_{2}})-(p_{1}.\epsilon_{p_{2}})(p_{2}.\epsilon_{p_{1}}) \Big],
\end{eqnarray}
where the $2!$ is due to the identical particles in the initial state. Notice that the second term vanishes since in $2 \rightarrow 2$ scattering, $\epsilon_{p_{1(2)}}$ is orthogonal to $p_{2(1)}$. On the other hand, the first term depends on the helicities of photons in the initial state. Inspecting eq. (\ref{eq:Pol_z}), we can see that if photons have the same (opposite) helicities, then $\epsilon_{p_{1}}.\epsilon_{p_{2}} = 1 (0)$. Therefore, the amplitude becomes
\begin{align}
\hat{\mathscr{M}}(\gamma_{\pm}\gamma_{\pm} \rightarrow W_{L}^{+}W_{L}^{-}) &= -4\Bigg( \frac{c^{\text{SM}}_{\gamma1}\delta_{\gamma 1}\alpha}{\pi v^{2}} \Bigg)  C_{i}C^{*}_{f} \int d\Phi_{4} \frac{E^{2}}{2} = - \frac{\sqrt{2}}{8}\Bigg( \frac{c^{\text{SM}}_{\gamma1}\delta_{\gamma 1}\alpha}{\pi^{2} v^{2}} \Bigg) E^{2},  \label{eq:SampleCalc2_1}\\
\hat{\mathscr{M}}(\gamma_{\pm}\gamma_{\mp} \rightarrow W_{L}^{+}W_{L}^{-}) &=  0. \label{eq:SampleCalc2_2}
\end{align}
\subsection{Unitarity from Partial Wave Expansion}\label{app:4.5}
In this paper, we used the unitarity condition given in eq. (\ref{eq:UnitarityBound}), which was derived in \cite{Chang:2019vez}. This approach is superior to the traditional approach that utilizes partial wave unitarity commonly used in the literature because it can be applied to any $m \rightarrow n$ scattering, whereas the latter approach can only be applied in the case of $2 \rightarrow 2$ scattering. In addition, even in $2 \rightarrow 2$ scattering, the condition in eq. (\ref{eq:UnitarityBound}) gives stronger bounds compared to those obtained from partial wave unitarity as we demonstrate below.

We begin by briefly reviewing partial wave unitarity following \cite{DiLuzio:2016sur}. The Lorentz-invariant amplitude $\mathscr{M}_{fi}$ can be expanded into partial waves of the total angular momentum $J$ \cite{Itzykson, Chanowitz:1978mv, Schuessler:2007av}, where the expansion coefficients are given by
\begin{equation}\label{eq:PartialWave}
a_{ij}^{J} = \frac{\beta_{i}^{1/4}(s,m_{i1}^{2},m_{i2}^{2})\beta_{f}^{1/4}(s,m_{f1}^{2},m_{f2}^{2})}{32 \pi s}\int_{-1}^{1} d\cos \theta d^{J}_{\mu_{i}\mu_{f}}(\theta) \mathscr{M}(\theta,\sqrt{s}),
\end{equation}
where $\beta(x,y,z) = x^{2} + y^{2} + z^{2} - 2x y -2 x z -2 y z$, $ d^{J}_{\mu_{i}\mu_{f}}$ is the $J$-th Wigner $d$-function in the Jacob-Wick expansion \cite{Jacob:1959at}, and for initial and final state particles with helicities $(\lambda_{i1},\lambda_{i2})$ and $(\lambda_{f1},\lambda_{f2})$, we have $\mu_{i} \equiv \lambda_{i2} - \lambda_{i1}$, $\mu_{f} \equiv \lambda_{f2} - \lambda_{f1}$. In eq. (\ref{eq:PartialWave}), an extra factor of $1/\sqrt{2}$ has to be included for identical initial or final state particles. For $\mu_{i}  = \mu_{f} =0$ the Wigner $d$-functions reduce to the Legendre polynomials; $d^{J}_{00} = P_{J}$, and in the massless limit $\beta(s,0,0) = s^{2}$. The unitarity of the $S$ matrix, $S^{\dagger}S = 1$, implies that
\begin{equation}\label{eq:Unitarity1}
\frac{1}{2i}\big(a^{J}_{fi} -a^{J*}_{if}) \geq \sum_{k} a^{J*}_{kf}a^{J}_{ki},
\end{equation}
where the sum goes over all 2-particle states. For $f=i$, eq. (\ref{eq:Unitarity1}) becomes
\begin{equation}\label{eq:Unitarity2}
\text	{Im}\hspace{1mm} a_{ii}^{J} \geq | a_{ii}^{J}|^{2},
\end{equation}
which implies that $ a_{ii}^{J}$ must lie inside the circle in the Argand plane defined by
\begin{equation}\label{eq:Unitarity3}
(\text{Re} \hspace{1mm}a_{ii}^{J})^{2} + \Big( \text	{Im}\hspace{1mm} a_{ii}^{J} - \frac{1}{2} \Big)^{2} \leq \frac{1}{4}.
\end{equation}
which implies
\begin{equation}\label{eq:Unitarity3}
|\text	{Im}\hspace{1mm} a_{ii}^{J}| \hspace{1mm} \leq 1, \hspace{1cm} |\text{Re} \hspace{1mm}a_{ii}^{J}| \hspace{1mm}\leq\frac{1}{2}.
\end{equation}

The second bound in eq. (\ref{eq:Unitarity3}) in particular, can be used for the expansion of tree-level amplitudes, and here we use it to find the partial wave unitarity bounds for the $2 \rightarrow 2$ processes studied in this paper, which are given in Tables \ref{tab:1} and \ref{tab:5}. 

For processes $\gamma\gamma \rightarrow X_{1}X_{2}$, $d^{J}_{ij} = d^{J}_{00} = P_{J}$, and only $P_{0}$ contributes. Explicit calculation of the bounds shows that they are weaker than the ones using eq. (\ref{eq:UnitarityBound}) by a factor of 2. On the other hand, for processes $\gamma  X_{1}\rightarrow \gamma X_{2}$, we need the function

\begin{equation}\label{eq:d11J}
d^{J}_{11}(\cos{\theta}) = \frac{(1+\cos{\theta})}{J(J+1)}\Big[P'_{J}(\cos{\theta}) - (1-\cos{\theta}) P''_{J}(\cos{\theta}) \Big],
\end{equation}
and one can show that the strongest bounds are obtained for $J=1,2$. Explicit calculation of the bounds shows that they are weaker than the ones using eq. (\ref{eq:UnitarityBound}) by a factor of 3. Thus, eq. (\ref{eq:UnitarityBound}) always yields stronger unitarity bounds.
\newpage
\subsection{Results}
\label{app:A5}
Here we present the leading high-energy behavior for the processes used in the main text. All massive gauge bosons are understood to be longitudinally polarized unless expressly indicated otherwise. We use $-(+)$ to denote LH (RH) photons/transverse modes. All processes of the form $\gamma_{\pm}\gamma_{\mp}X \rightarrow Y$ or $\gamma_{\pm} X \rightarrow \gamma_{\pm} Y$ have vanishing amplitudes as they do not conserve angular momentum. All other processes not listed in the tables are related to the ones listed in the tables via charge conjugation. All amplitudes are calculated	 in the contact approximation, and all particles are assumed to be massless.

\begin{table}[!ht]
\centerline{
\begin{minipage}{0.8\textwidth}
\centering
\vspace{1 mm}
\tabcolsep3pt\begin{tabular}{|c|c||c|c|}
\hline
\textbf{Process} & \textbf{$\times \alpha\frac{c^{\text{SM}}_{\gamma 1} |\delta_{\gamma 1}|}{8\pi^{2}v^{2}}E^{2}$} &  \textbf{Process} & \textbf{$\times\alpha\frac{c^{\text{SM}}_{\gamma 1} |\delta_{\gamma 1}|}{8\pi^{2}v^{2}}E^{2}$} \\
\hline
$\gamma_{\pm}\gamma_{\pm} \rightarrow W^{+}W^{-}$ & $\sqrt{2}$ & $\gamma_{\pm} W^{+} \rightarrow \gamma_{\mp} W^{+}$ & $1$\\ 
$\gamma_{\pm}\gamma_{\pm} \rightarrow ZZ$ & $1$ & $\gamma_{\pm} Z\rightarrow \gamma_{\mp} Z$ & $1$\\ 
\hline
\textbf{Process} & \textbf{$\times\alpha \frac{c^{\text{SM}}_{\gamma 1} |\delta_{\gamma 2}|}{8\pi^{2}v^{2}}E^{2}$} &  \textbf{Process} & \textbf{{$\times \alpha\frac{c^{\text{SM}}_{\gamma 1} |\delta_{\gamma 2}|}{8\pi^{2}v^{2}}E^{2}$}} \\
\hline
$\gamma_{\pm}\gamma_{\pm} \rightarrow hh$ & $1$ & $\gamma_{\pm} h\rightarrow \gamma_{\mp} h$ & $1$\\ 
\hline
\end{tabular}
\caption{\label{tab:1} \small $|\hat{\mathscr{M}}|$ of the 4-body model-independent unitarity-violating processes arising from the modification of the Higgs coupling to $\gamma\gamma$. }
\end{minipage}}
\end{table}

\begin{table}[!ht]
\centerline{
\begin{minipage}{0.8\textwidth}
\centering
\vspace{1 mm}
\tabcolsep3pt\begin{tabular}{|c|c||c|c|}
\hline
\textbf{Process} & \textbf{$\times \alpha \frac{c^{\text{SM}}_{\gamma 1} |\delta_{\gamma 2} - \delta_{\gamma 1}|}{96 \sqrt{2}\pi^{3}v^{3}} E^{3}$} &  \textbf{Process} & \textbf{$\times \alpha \frac{c^{\text{SM}}_{\gamma 1} |\delta_{\gamma 2} - \delta_{\gamma 1}|}{96 \sqrt{2}\pi^{3}v^{3}} E^{3}$} \\
\hline
$\gamma_{\pm}\gamma_{\pm} \rightarrow hZ^{2}$ & $3$ & $\gamma_{\pm}\gamma_{\pm} h\rightarrow Z^{2} $ & $1$\\ 
$\gamma_{\pm}\gamma_{\pm} Z \rightarrow hZ$ & $\sqrt{2}$ & $\gamma_{\pm} Z\rightarrow \gamma_{\mp} h Z $ & $2$\\ 
$\gamma_{\pm}h \rightarrow \gamma_{\mp} Z^{2}$ & $\sqrt{2}$ & $\gamma_{\pm}\gamma_{\pm} \rightarrow hW^{+}W^{-}  $ & $3\sqrt{2}$\\ 
$\gamma_{\pm}\gamma_{\pm}h \rightarrow W^{+}W^{-}$ & $\sqrt{2}$ & $\gamma_{\pm}\gamma_{\pm} W^{+}\rightarrow h W^{+}  $ & $\sqrt{2}$\\ 
$\gamma_{\pm}h \rightarrow \gamma_{\mp} W^{+}W^{-}$ & $2$ & $\gamma_{\pm}W^{+}\rightarrow \gamma_{\mp}  h W^{+}  $ & $2$\\ 
\hline
\textbf{Process} & \textbf{$\times \alpha \frac{ c^{\text{SM}}_{\gamma 1} |\delta_{\gamma 3}| }{96 \sqrt{2}\pi^{3}v^{3}} E^{3}$} & \textbf{Process} & \textbf{$\times \alpha \frac{ c^{\text{SM}}_{\gamma 1} |\delta_{\gamma 3}|}{96 \sqrt{2}\pi^{3}v^{3}} E^{3}$} \\
\hline
$\gamma_{\pm}\gamma_{\pm} \rightarrow h^{3}$ & $\sqrt{3}$ &
$\gamma_{\pm}\gamma_{\pm} h \rightarrow h^{2}$ & $1$\\ 
$\gamma_{\pm} h \rightarrow \gamma_{\mp} h^{2}$ & $\sqrt{2}$ & &\\ 
\hline
\end{tabular}
\caption{\label{tab:2} \small $|\hat{\mathscr{M}}|$ of the 5-body unitarity-violating processes arising from the modification of the Higgs coupling to $\gamma\gamma$.
}
\end{minipage}}
\end{table}

\begin{table}[!ht]
\centerline{
\begin{minipage}{0.8\textwidth}
\centering
\vspace{1 mm}
\tabcolsep3pt\begin{tabular}{|c|c|}
\hline
\textbf{Process} & \textbf{$\times \alpha \frac{ c^{\text{SM}}_{\gamma 1} |\delta_{\gamma 2} - \delta_{\gamma 1}|}{768 \pi^{4}v^{4}} E^{4}$} \\
\hline
$\gamma_{\pm} Z^{2}\rightarrow \gamma_{\mp} Z^{2}$ & $2$ \\
$\gamma_{\pm}\gamma_{\pm} Z\rightarrow Z^{3} $ & $\sqrt{3}$\\ 
$\gamma_{\pm} ZW^{+}\rightarrow \gamma_{\mp} ZW^{+}$ & $\frac{4}{3}$ \\
$\gamma_{\pm} W^{+}W^{-} \rightarrow \gamma_{\mp}Z^{2} $ & $\frac{2\sqrt{2}}{3}$\\
$\gamma_{\pm} \gamma_{\pm} Z \rightarrow ZW^{+}W^{-} $ & $\sqrt{2}$\\
$\gamma_{\pm} \gamma_{\pm}W^{+}\rightarrow Z^{2}W^{+}$ & $1$ \\
$\gamma_{\pm} W^{+}W^{-} \rightarrow \gamma_{\mp} W^{+}W^{-} $ & $\frac{8}{3}$\\
$ \gamma_{\pm} W^{+}W^{+} \rightarrow \gamma_{\mp} W^{+}W^{+} $ & $\frac{4}{3}$ \\
 $\gamma_{\pm} \gamma_{\pm}W^{+}\rightarrow W^{+}W^{-}W^{+}$ & $2$ \\
\hline
\textbf{Process} & \textbf{$\times \alpha \frac{ c^{\text{SM}}_{\gamma 1}|\delta_{\gamma 1} - \delta_{\gamma 2}+\frac{1}{2}\delta_{\gamma 3}|}{576 \pi^{4}v^{4}} E^{4}$} \\
\hline
$\gamma_{\pm} hW^{+}\rightarrow \gamma_{\mp} hW^{+}$ & $2$ \\
$\gamma_{\pm} h^{2}\rightarrow \gamma_{\mp} W^{+}W^{-}$ & $\sqrt{2}$ \\
$\gamma_{\pm} \gamma_{\pm} h\rightarrow h W^{+}W^{-}$ & $\frac{3}{\sqrt{2}}$ \\
$\gamma_{\pm} \gamma_{\pm} W^{+}\rightarrow h^{2} W^{+}$ & $\frac{3}{2}$ \\
$\gamma_{\pm} hZ \rightarrow \gamma_{\mp} h Z$ & $2$ \\
$\gamma_{\pm} h^{2} \rightarrow \gamma_{\mp} Z^{2}$ & $1$ \\
$\gamma_{\pm} \gamma_{\pm} h\rightarrow h Z^{2}$ & $\frac{3}{2}$ \\
$\gamma_{\pm} \gamma_{\pm} Z\rightarrow h^{2} Z$ & $\frac{3}{2}$ \\
\hline
\textbf{Process} & \textbf{$\times \alpha \frac{ c^{\text{SM}}_{\gamma 1} |\delta_{\gamma 4}|}{1152 \pi^{4}v^{4}} E^{4}$} \\
\hline
$\gamma_{\pm} h^{2}\rightarrow \gamma_{\mp} h^{2}$ & $1$ \\
$\gamma_{\pm} \gamma_{\pm} h\rightarrow h^{3}$ & $\frac{\sqrt{3}}{2}$ \\
\hline
\end{tabular}
\caption{\label{tab:3} \small $|\hat{\mathscr{M}}|$ of the 6-body unitarity-violating processes arising from the modification of the Higgs coupling to $\gamma\gamma$.
}
\end{minipage}}
\end{table}

\begin{table}[!ht]
\centerline{
\begin{minipage}{0.8\textwidth}
\centering
\vspace{1 mm}
\tabcolsep3pt\begin{tabular}{|c|c|}
\hline
\textbf{Process} & \textbf{$\times 4 \alpha c^{\text{SM}}_{\gamma 1} |\delta_{\gamma 2n}| \Big( \frac{E}{4\pi v}\Big)^{2n}$} \\
\hline
$\gamma_{\pm} \gamma_{\pm}  h^{n-1}\rightarrow h^{n+1}$ & $\frac{\sqrt{2}}{(n+1)!n!\sqrt{(n+1)!(n-1)!}}$ \\
$\gamma_{\pm} h^{n}\rightarrow  \gamma_{\mp} h^{n}$ & $\frac{2}{(n+1)!(n+1)!(n-1)!}$ \\
\hline
\end{tabular}
\caption{\label{tab:4} \small $|\hat{\mathscr{M}}|$ of the $2n+2$-body model-independent unitarity-violating processes arising from the modification of the Higgs coupling to $\gamma\gamma$.}
\end{minipage}}
\end{table}

\begin{table}[!ht]
\centerline{
\begin{minipage}{0.8\textwidth}
\centering
\vspace{1 mm}
\tabcolsep3pt\begin{tabular}{|c|c||c|c|}
\hline
\textbf{Process} & \textbf{$\times \alpha \frac{c^{\text{SM}}_{\gamma Z1}|\delta_{\gamma Z1}| }{16\pi^{2}v^{2}}E^{2}$} &  \textbf{Process} & \textbf{$\times \alpha \frac{c^{\text{SM}}_{\gamma Z1}|\delta_{\gamma Z1}|}{16\pi^{2}v^{2}}E^{2}$} \\
\hline
$\gamma_{\pm} Z_{T\pm} \rightarrow W^{+}W^{-}$ & $2$ & 
$\gamma_{\pm} W^{+} \rightarrow Z_{T\mp} W^{+}$ & $1$\\ 
$\gamma_{\pm} Z_{T\pm} \rightarrow ZZ$ & $\sqrt{2}$ & $\gamma_{\pm} Z\rightarrow Z_{T\mp} Z$ & $1$\\ 
\hline
\textbf{Process} & \textbf{$\times \alpha \frac{c^{\text{SM}}_{\gamma Z1} |\delta_{\gamma Z2}|}{16\pi^{2}v^{2}}E^{2}$} &  \textbf{Process} & \textbf{$\times\alpha \frac{c^{\text{SM}}_{\gamma Z1} |\delta_{\gamma Z2}|}{16\pi^{2}v^{2}}E^{2}$} \\
\hline
$\gamma_{\pm}Z_{T\pm} \rightarrow hh$ & $\sqrt{2}$ & $\gamma_{\pm} h\rightarrow Z_{T\mp} h$ & $1$\\ 
\hline
\end{tabular}
\caption{\label{tab:5} \small $|\hat{\mathscr{M}}|$ of the 4-body model-independent unitarity-violating processes arising from the modification of the Higgs coupling to $\gamma Z$. }
\end{minipage}}
\end{table}

\begin{table}[!ht]
\centerline{
\begin{minipage}{0.8\textwidth}
\centering
\vspace{1 mm}
\tabcolsep3pt\begin{tabular}{|c|c||c|c|}
\hline
\textbf{Process} & \textbf{$\times \alpha \frac{ c^{\text{SM}}_{\gamma Z1} |\delta_{\gamma Z2} - \delta_{\gamma Z1}| }{192 \pi^{3}v^{3}} E^{3}$} &  \textbf{Process} & \textbf{$\times \alpha \frac{ c^{\text{SM}}_{\gamma Z1} |\delta_{\gamma Z2} - \delta_{\gamma Z1}|}{192 \pi^{3}v^{3}} E^{3}$} \\
\hline
$\gamma_{\pm} Z_{T\pm} \rightarrow hZ^{2}$ & $3$ & $\gamma_{\pm}Z_{T\pm} h\rightarrow Z^{2} $ & $1$\\ 
$\gamma_{\pm}Z_{T\pm} Z \rightarrow hZ$ & $\sqrt{2}$ & $\gamma_{\pm} Z\rightarrow Z_{T\mp} h Z $ & $\sqrt{2}$\\ 
$Z_{T\pm}Z \rightarrow \gamma_{\mp} Z h$ & $\sqrt{2}$ & $\gamma_{\pm}h \rightarrow Z_{T\mp} Z^{2}  $ & $1$\\ 
$Z_{T\pm} h \rightarrow \gamma_{\mp} Z^{2}$ & $1$ & $\gamma_{\pm}Z_{T\pm} \rightarrow h W^{+}W^{-}  $ & $3\sqrt{2}$\\ 
$\gamma_{\pm}Z_{T\pm}h \rightarrow W^{+}W^{-}$ & $\sqrt{2}$ & $\gamma_{\pm}Z_{T\pm}W^{+}\rightarrow  h W^{+}  $ & $\sqrt{2}$\\ 
$\gamma_{\pm}h \rightarrow Z_{T\mp} W^{+}W^{-}$ & $\sqrt{2}$ & $Z_{T\pm} h \rightarrow \gamma_{\mp} W^{+}W^{-}$ & $\sqrt{2}$\\
$\gamma_{\pm}W^{+} \rightarrow Z_{T\mp} hW^{+}$ & $\sqrt{2}$ & $Z_{T\pm} W^{+} \rightarrow \gamma_{\mp} h W^{+}$ & $\sqrt{2}$\\
\hline
\textbf{Process} & \textbf{$\times \alpha \frac{c^{\text{SM}}_{\gamma Z1} |\delta_{\gamma Z3}|  }{192\pi^{3}v^{3}} E^{3}$} & \textbf{Process} & \textbf{$\times \alpha \frac{c^{\text{SM}}_{\gamma Z1} |\delta_{\gamma Z3}| }{192\pi^{3}v^{3}} E^{3}$} \\
\hline
$\gamma_{\pm}Z_{T\pm} \rightarrow h^{3}$ & $\sqrt{3}$&
$\gamma_{\pm}Z_{T\pm} h \rightarrow h^{2}$ & $1$\\
$\gamma_{\pm} h \rightarrow Z_{T\mp} h^{2}$ & $1$ &
$Z_{T\pm} h \rightarrow \gamma_{\mp} h^{2}$ & $1$\\ 
\hline
\end{tabular}
\caption{\label{tab:6} \small $|\hat{\mathscr{M}}|$ of the 5-body unitarity-violating processes arising from the modification of the Higgs coupling to $\gamma Z$.
}
\end{minipage}}
\end{table}

\begin{table}[!ht]
\centerline{
\begin{minipage}{0.8\textwidth}
\centering
\vspace{1 mm}
\tabcolsep7pt\begin{tabular}{|c|c|}
\hline
\textbf{Process} & \textbf{$\times \alpha \frac{ c^{\text{SM}}_{\gamma Z1} |\delta_{\gamma Z2} - \delta_{\gamma Z1}|}{1152\sqrt{2} \pi^{4}v^{4}} E^{4}$} \\
\hline
$\gamma_{\pm} Z^{2}\rightarrow Z_{T\mp} Z^{2}$ & $\frac{3}{\sqrt{2}}$ \\
$\gamma_{\pm}Z_{T\pm} Z\rightarrow Z^{3} $ & $\frac{3\sqrt{3}}{2}$\\ 
$\gamma_{\pm} ZW^{+}\rightarrow Z_{T\mp} ZW^{+}$ & $\sqrt{2}$ \\
$\gamma_{\pm} W^{+}W^{-} \rightarrow Z_{T\mp}Z^{2} $ & $1$\\
$Z_{T\pm} W^{+}W^{-} \rightarrow \gamma_{\mp} Z^{2} $ & $1$\\
$\gamma_{\pm} Z_{T\pm}Z\rightarrow ZW^{+}W^{-}$ & $\frac{3}{\sqrt{2}}$ \\
$\gamma_{\pm} Z_{T\pm} W^{+} \rightarrow Z^{2}W^{+} $ & $\frac{3}{2}$\\
$ \gamma_{\pm} W^{+}W^{-} \rightarrow Z_{T\mp} W^{+}W^{-} $ & $2\sqrt{2}$ \\
$ \gamma_{\pm} W^{+}W^{+} \rightarrow Z_{T\mp} W^{+}W^{+} $ & $\sqrt{2}$ \\
$\gamma_{\pm} Z_{T\pm} W^{+} \rightarrow W^{+}W^{-}W^{+} $ & $3$\\
\hline
\textbf{Process} & \textbf{$\times \alpha \frac{c^{\text{SM}}_{\gamma Z1}|\delta_{\gamma Z1} - \delta_{\gamma Z2}+\frac{1}{2}\delta_{\gamma Z3}|}{1152 \pi^{4}v^{4}} E^{4}$} \\
\hline
$\gamma_{\pm} hW^{+}\rightarrow Z_{T\mp} hW^{+}$ & $2$ \\
$\gamma_{\pm} h^{2}\rightarrow Z_{T\mp} W^{+}W^{-}$ & $\sqrt{2}$ \\
$\gamma_{\pm} W^{+}W^{-} \rightarrow Z_{T\mp} h^{2}$ & $\sqrt{2}$ \\
$\gamma_{\pm} Z_{T\pm} h\rightarrow h W^{+}W^{-}$ & $3$ \\
$\gamma_{\pm} Z_{T\pm} W^{+} \rightarrow h^{2} W^{+}$ & $\frac{3}{\sqrt{2}}$ \\
$\gamma_{\pm} hZ \rightarrow Z_{T\mp} hZ$ & $2$ \\
$\gamma_{\pm} h^{2}\rightarrow Z_{T\mp} Z^{2}$ & $1$ \\
$\gamma_{\pm} Z^{2}\rightarrow Z_{T\mp} h^{2}$ & $1$ \\
$\gamma_{\pm} Z_{T\pm} h\rightarrow h Z^{2}$ & $\frac{3}{\sqrt{2}}$ \\
$\gamma_{\pm} Z_{T\pm} Z\rightarrow Z h^{2}$ & $\frac{3}{\sqrt{2}}$ \\
\hline
\textbf{Process} & \textbf{$\times \alpha \frac{ c^{\text{SM}}_{\gamma Z1} |\delta_{\gamma Z4}|}{2304 \pi^{4}v^{4}} E^{4}$} \\
\hline
$\gamma_{\pm} h^{2}\rightarrow Z_{T\mp} h^{2}$ & $1$ \\
$\gamma_{\pm} Z_{T\pm} h\rightarrow h^{3}$ & $\sqrt{\frac{3}{2}}$ \\
\hline
\end{tabular}
\caption{\label{tab:8} \small $|\hat{\mathscr{M}}|$ of the 6-body unitarity-violating processes arising from the modification of the Higgs coupling to $\gamma Z$.
}
\end{minipage}}
\end{table}

\clearpage

\begin{table}[!ht]
\centerline{
\begin{minipage}{0.8\textwidth}
\centering
\vspace{1 mm}
\tabcolsep7pt\begin{tabular}{|c|c|}
\hline
\textbf{Process} & \textbf{$\times 4 \alpha c^{\text{SM}}_{\gamma Z1} |\delta_{\gamma Z2n}| \Big( \frac{E}{4\pi v}\Big)^{2n}$} \\
\hline
$\gamma_{\pm} Z_{T\pm}  h^{n-1}\rightarrow h^{n+1}$ & $\frac{1}{(n+1)!n!\sqrt{(n+1)!(n-1)!}}$ \\
$\gamma_{\pm} h^{n}\rightarrow  Z_{T\mp} h^{n}$ & $\frac{1}{(n+1)!(n+1)!(n-1)!}$ \\
\hline
\end{tabular}
\caption{\label{tab:11} \small $|\hat{\mathscr{M}}|$ of the $2n+2$-body model-independent unitarity-violating processes arising from the modification of the Higgs coupling to $\gamma Z$.}
\end{minipage}}
\end{table}

\end{document}